\pdfoutput=1

\documentclass[11pt]{article}

\usepackage{amsmath}
\usepackage{enumerate}
\usepackage{amsthm}
\usepackage{eucal}
\usepackage{amssymb}
\usepackage{mathrsfs}
\usepackage[final]{graphicx}
\usepackage{multirow}
\usepackage{listings}
\usepackage{mathrsfs}
\usepackage{color}

\usepackage{grffile}
\usepackage{xcolor}
\usepackage{subcaption}

\usepackage{authblk}
\usepackage{soul}

\newcommand{\argmin}{\arg\!\min}

\begin{document}
\author[1]{Iris Ivy Gauran}
\author[1]{Junyong Park}
\author[2]{Johan Lim}
\author[1]{DoHwan Park}
\author[1]{John Zylstra}
\author[3]{Thomas Peterson}
\author[3]{Maricel Kann}
\author[4]{John Spouge}

\affil[1]{Department of Mathematics and Statistics, University of Maryland, Baltimore County, Baltimore, MD 21250, USA}
\affil[2]{Department of Statistics, Seoul National University, Seoul, 08826, Republic of Korea}
\affil[3]{Department of Biological Sciences, University of Maryland, Baltimore County, Baltimore, MD 21250, USA}
\affil[4]{National Center for Biotechnology Information, National Library of Medicine, National Institutes of Health, Bethesda, MD 20894, USA}

\title{\textbf{Empirical Null Estimation using Discrete Mixture Distributions
		and its Application to\\
		Protein Domain Data}}
\date{}
\maketitle
\begin{abstract}
	\noindent In recent mutation studies, analyses based on protein domain positions are gaining popularity over gene-centric approaches since the latter have limitations in considering the functional context that the position of the mutation provides.  This presents a large-scale simultaneous inference problem, with hundreds of hypothesis tests to consider at the same time.  This paper aims to select significant mutation counts while controlling a given level of Type I error via False Discovery Rate (FDR) procedures.  One main assumption is that there exists a cut-off value such that smaller counts than this value are generated from the null distribution.  We present several data-dependent methods to determine the cut-off value. We also consider a two-stage procedure based on screening process so that the number of mutations exceeding a certain value should be considered as significant mutations. Simulated and protein domain data sets are used to illustrate this procedure in estimation of the empirical null using a mixture of discrete distributions.
\end{abstract}

\vspace{5mm}
\section{Introduction}
\vspace{3mm}
\noindent Interest towards multiple testing procedures has been growing rapidly in the advent of the so-called genomic age.  With the breakthrough in large-scale methods to purify, identify and characterize DNA, RNA, proteins and other molecules, researchers are becoming increasingly reliant on statistical methods for determining the significance of biological findings (\cite{Pollard2005test}).  Gene-based analyses of cancer data are classic examples of studies which present thousands of genes for simultaneous hypothesis testing.  However, \cite{Nehrt2012domain} reported that gene-centric cancer studies are limited since the functional context that the position of the mutation provides is not considered.  In lieu of this, \cite{Nehrt2012domain} and \cite{Yang2015protein} have shown that protein domain level analyses of cancer somatic variants could provide additional insights.  \\
\newpage
\noindent
In particular, these studies can identify functionally relevant somatic mutations where traditional gene-centric methods fail by focusing on protein domain regions within genes, leveraging the modularity and polyfunctionality of genes.  In protein domain-centric analyses of somatic mutations, somatic mutations from sequenced tumor samples are mapped from their genomic positions to positions within protein domains, enabling the comparison of distant genomic regions that share similar structure and amino acid composition (\cite{Peterson2010dmdm}; \cite{Peterson2012incorporating}; \cite{Peterson2013protein}).
In the analysis of sequenced tumor samples, it is assumed that the mutational distribution will consist of many ``passenger" mutations, which are non-functional randomly distributed background mutations, in addition to rare functional ``driver" mutations that reoccur at specific sites within the domain and contribute to the initiation or progression of cancer (\cite{Parmigiani2007statistical}; \cite{Stratton2009cancer}; \cite{Stratton2011exploring}). The problem that is addressed here is in a single domain, how to identify the highly mutated positions compared to the background where the number of positions in a domain can be as large as several tens or hundreds.
\\\\
\noindent
Motivated by the aforementioned domain-level analyses, we propose a methodology for identifying significant mutation counts while controlling the rate of false rejections.  \cite{Dudoit2003multiple} reported that much of the statistics microarray literature is focused on controlling the probability of a Type I error, a ``false discovery".  A traditional approach is to control the family-wise error rate (FWER), the probability of making at least one false discovery.  However, with the collection of simultaneous hypothesis tests in the hundreds or thousands, trying to limit the probability of even a single false discovery leads to lack of power.  Alternatively, in a seminal paper, \cite{BenjaminiHochberg} introduced a multiple-hypothesis testing error measure called False Discovery Rate (FDR).  This quantity is the expected proportion of false positive findings among all the rejected hypotheses.  Among the FDR-controlling test methods, \cite{Efron2001empirical} developed an empirical Bayes approach where they established a close connection between the estimated posterior probabilities and a local version of the FDR.\\
\\
\noindent A key step in controlling the local false discoveries is to estimate the null distribution of the test statistics. \cite{Efron2004large} stated that the test statistics in large-scale testing may not accurately follow the theoretical null distribution. Instead, the density of the null distribution is estimated from the large number of genes.  In these microarray experiments, \cite{Efron2005local} employed a normal mixture model and proposed maximum likelihood and mode matching to estimate the empirical null distribution. Using the same normal mixture model, \cite{JinCai2007} proposed a method to estimate the empirical null based on characteristic functions.  In addition, \cite{Park2011estimation} proposed a local FDR estimation procedure based on modeling the null distribution with a mixture of normal distributions.  However, these existing methods are based on the assumption that the null is a mixture of continuous distributions.  In the case of domain-level analyses, the data is characterized as mutation counts among $N$ positions in the domain.  This indicates that the available methods in the estimation of the empirical null should be extended to a mixture of discrete distributions.\\
\\
The rest of the paper is organized as follows. In Section 2, we discuss the problem in detail and review two existing multiple testing procedures, namely Efron's Local FDR procedure and Storey's procedure.  In Section 3, we introduce the estimation procedure for $f_0, f$ and $\pi_0$, where the null distribution is assumed to be a zero-inflated model. Also, a novel two-stage multiple testing procedure is presented in this section.  In Section 4, the performance of the new procedure is studied via simulations and the results for real data sets are presented.  Some concluding remarks will be presented in Section 5.

\vspace{5mm}
\section{Multiple Testing Procedures controlling FDR}
\vspace{3mm}
\noindent In this section, we briefly discuss the motivating example and review the existing procedures for analysis.  The collection of the original dataset is  $\boldsymbol{a} = (a_1, a_2, \ldots , a_N)'$, where $a_i$ is the number of mutations in the \emph{i}th position of the specific domain with $N$ positions. We define $\cal A$ = $\{j: j \geq 0, n_j > 0\}$ as the set of the unique values of $\boldsymbol{a}$, $K = \max (\boldsymbol{a})$, and $L$ is the cardinality of $\cal A$ where $L \leq K +1$.  Some relevant features of $\boldsymbol{a}$ follow.  A large proportion of positions do not have any mutation, $a_i = 0$. Also, $L$ is relatively small compared to $N$, which means that the number of mutations in many positions are tied. Since our goal is to identify the positions with extra disease mutation counts, it is only reasonable to have the same conclusion for positions wherein the number of mutations are tied. Therefore, we transform the data into the observed ``histograph" of positions over ``mutation counts". We define $n_j = \mid \{i: a_i = j\} \mid$, as the number of positions with $j$ mutations, $j \in \cal A$, and $\sum \limits_{j = 0}^K n_j = N$.  The ordered data $\boldsymbol{x}_N$ can be represented as a partition of the unique values of $\boldsymbol{a}$, that is,
$$\boldsymbol{x}'_N = (\boldsymbol{x}'_0, \boldsymbol{x}'_1, \ldots, \boldsymbol{x}'_K) = (\underbrace{0, 0, \ldots, 0}_{\boldsymbol{x}'_0}, \underbrace{1, 1, \ldots, 1}_{\boldsymbol{x}'_1} \ldots, \underbrace{K, K, \ldots, K}_{\boldsymbol{x}'_K})$$ where $\boldsymbol{x}_j$ is the column vector containing $n_j$ of $j$s.  Since the information contained in $\boldsymbol{x}_j$ is analogous to knowing $n_j$, for any $j \in \cal A$, then another, equivalent format of the data set is $\boldsymbol{y}_N = (n_0, n_1, \ldots, n_K)'$.\\
\\
\noindent For any single domain of interest, a total of $L$ mutation counts can be decomposed into two groups, $\mathcal{A}_0$ and $\mathcal{A}_1$, where $\mathcal{A}_0$ is the collection of small number of mutation counts which is considered to be non-significant and $\mathcal{A}_1$ is the set of large number of mutation counts which consists of significantly mutated positions.  Let the prior probabilities of the two groups be $\pi_0$ or $\pi_1 = 1 - \pi_0$, and assume corresponding densities, $f_0$ or $f_1$.  Define $f_0$ to be the null distribution and $f_1$ to be the alternative distribution. Therefore, we consider the problem of testing $L$ null hypotheses simultaneously,
$$ H_0:  \mbox{ $H_{0j}$ is true for $j \in {\cal A}$} $$
on the basis of a data set  $\boldsymbol{a}$, where $H_{0j}$ is stated as the number of mutations $j$ is generated from $f_0$
for all $j \in \mathcal{A}$ with $|\mathcal{A}| = L$. For a given position, the number of mutations follow one of the two distributions $f_0$ or $f_1$, so the probability density function of the mixture distribution can be represented as
\begin{eqnarray}\label{eqn:f}
	f(a) = \pi_0 f_0 (a) + (1 - \pi_0) f_1 (a)
\end{eqnarray}
\noindent and our goal is to identify the positions which have significantly different patterns from the null.\\
\\
For continuous data, \cite{Efron2005local} introduced the idea of ``zero assumption" where observations around the central peak of the distribution consists mainly of null cases. Using this assumption, $f_0$ is estimated using Gaussian quadrature which is based on derivative at the mode.  However,  such a procedure is not applicable to discrete data.
In our problem on discrete data, we introduce the following assumption on the null distribution which plays a key role throughout this paper.
\newpage
\noindent
\subsection*{Assumption on $f_0$:}
\begin{equation} \label{eqn:assumption}
f_1 (a) =0 \hspace{2mm}for\hspace{2mm} a \leq C \hspace{2mm}for\hspace{2mm} some\hspace{2mm} C \in \mathbb{Z}^+.
\end{equation}
\noindent
From the assumption, $a_i \leq C$  are guaranteed to be from $f_0$ and
$a_i > C$ are generated from the mixture of $f_0$ and $f_1$.  For notational convenience, we relabel the data as $\boldsymbol{x}_n = (\boldsymbol{x}_0, \boldsymbol{x}_1, \ldots, \boldsymbol{x}_C)$ for the null sample and $\boldsymbol{x}_{N - n} = (\boldsymbol{x}_{C + 1}, \boldsymbol{x}_{C + 2}, \ldots, \boldsymbol{x}_K)$ for the mixture of null and non-null samples.  The sampling distribution for the null sample is $f_0$ itself while $f$ in \eqref{eqn:f} is the sampling distribution of the non-null sample. We will discuss more details about how to choose the value of $C$ in the next section.
\\\\
Following the pioneering work of \cite{BenjaminiHochberg}, we employ the sequential p-value method to determine $r$ that tells us to reject $p_{(1)}, p_{(2)}, \ldots, p_{(r)}$, where $p_{(1)} , p_{(2)}, \ldots, p_{(K)}$ are the ordered observed p-values.  \cite{Storey2002direct} improved the Benjamini-Hochberg procedure with the inclusion of the estimator of the null proportion, $\hat{\pi}_0$, which indicates that we reject $p_{(1)}, p_{(2)}, \ldots, p_{(l)}$ such that
\begin{eqnarray}
l = max \left\{i: p_{(i)} \leq \displaystyle \frac{\alpha \displaystyle \sum \limits_{j \geq i} n_j}{N\hat{\pi}_0} \right \}
\end{eqnarray}
The BH procedure and Storey's procedure are equivalent, that is $r = l$, if we take $\hat{\pi}_0 = 1$.
The details about the estimation of $\pi_0$ is provided in the next section.  Moreover, following \cite{Efron2012large}, we define the local FDR at any mutation count, say $t$, as
\begin{equation}\label{eqn:lfdr}
fdr(t) = \displaystyle \frac{\pi_0 f_0 (t)}{f(t)}
\end{equation}
which indicates that $fdr(t)$ is the posterior probability of a true null hypothesis at $t$. The interpretation of the local FDR value is analogous to the frequentist's p-value wherein local FDR values less than a specified level of significance provide stronger evidence against the null hypothesis.\\

\vspace{5mm}
\section{Methodology}
\subsection{Model Specification}
\vspace{3mm}
Depending on the application, we assume that the mutation counts follow a zero-inflated model in order to account for the true zeros in the count model and the excess zeros.  The class of models considered is the Generalized Poisson (GP) distribution introduced by \cite{Consul1970generalization}, with an additional zero-inflation parameter.  \\
\\
Let $T$ be a nonnegative integer-valued random variable where relative to Poisson model, it is overdispersed with variance to mean ratio exceeding 1. If $T \sim GP(\lambda, \theta)$, then the probability mass function can be written as
\begin{eqnarray}
	P(T = t) = g (t) = \displaystyle \frac{\lambda(\lambda + \theta t)^{t - 1}}{t!}e^{-\lambda - \theta t}
\end{eqnarray}
where $0 \leq \theta < 1$ and $\lambda > 0$.
\newpage
\noindent
If zero is observed with a significantly higher frequency, we can include a zero-inflation parameter to characterize the distribution.  Then $X \sim ZIGP(\eta, \lambda, \theta)$ and the probability that $X = j$, denoted by $f_0 (j)$, is
\[f_0 (j) = \begin{cases}
\eta + (1 - \eta)e^{-\lambda} & j = 0 \\
(1 - \eta)g (j) & j = 1, 2, \ldots
\end{cases}
\]
where $j$ is a nonnegative integer, $0 \leq \eta < 1, \hspace{1mm}0 \leq \theta < 1$ and $\lambda > 0$.  Recently, ZIGP models have been found useful for the analysis of heavy-tailed count data with a large proportion of zeros (\cite{Gupta2005score}; \cite{Famoye2006zero}; \cite{Gschlossl2008modelling}).  The ZIGP model reduces to Zero-Inflated Poisson (ZIP) distribution when $\theta = 0$, Generalized Poisson distribution (GP) when $\eta = 0$ and Poisson distribution when $\eta = 0$ and $\theta = 0$.\\
\\
The ZIP model, first introduced by \cite{Lambert1992zero}, is applied when the count data possess the equality of mean and
variance property while taking into consideration the structural zeros and zeros which exist by chance.  Meanwhile, the Zero-Inflated Negative Binomial (ZINB) model is widely used for handling data with population heterogeneity which may be caused by the occurrence of excess zeros and the overdispersion due to
unobserved heterogeneity (\cite{Phang2013zero}). Several studies show that ZINB model provides a better fit to the overdispersed count data when ZIP is inadequate (\cite{Xia2012modeling}; \cite{Ullah2010statistical}; \cite{Kibria2006applications}).
However, \cite{Joe2005generalized} showed that the ZIGP distribution provides a better fit than ZINB when there is a large fraction of zeros and the data is heavily right-skewed.  They compared the probabilistic properties of the zero-inflated variations of NB and GP distributions, such as probability mass and skewness, while keeping the first two moments fixed.  Using this result, it is worthwhile to consider ZIGP rather than ZINB given that the mutation count data exhibited both features.

\vspace{5mm}
\subsection{Estimation of $f_0, f$ and $\pi_0$}
\vspace{3mm}
From (\ref{eqn:lfdr}), the local FDR formulation consists of unknown quantities $f_0$, $f$, and $\pi_0$ which must be estimated accordingly. We follow the idea of ``zero assumption" in \cite{Efron2005local} which modeled  $f_0$ to normal null and \cite{Park2011estimation} which modeled $f_0$ as a mixture of normals. In the proposed method, we apply $f_0$ to the context of ZIP and ZIGP models which indicate that a small mutation count suggests a few random background mutations, whereas a large mutation count suggests a mixture of a few background and a lot of functional disease mutations.  However, since $f_0$ is unknown in practice, four count models will be compared in order to come up with estimates for the parameters of the null distribution.  These models belong to the class of ZIGP distribution, namely, (1) ZIGP (2) ZIP (3) Generalized Poisson and (4) Poisson.  If the true $f_0$ is ZIGP and the model used to estimate $f_0$ is ZIGP then we expect superior results compared to the other three distributions.  Moreover, if the true null distribution is ZIP, then we expect better results for ZIP and ZIGP distribution compared to GP and Poisson distribution.  This suggests that since ZIGP can characterize overdispersion, even if there is none such as the case of ZIP, it should still be able to capture the behavior of $f_0$ accurately.  \\\\
\noindent To estimate the parameters of $f_0$ for any of these four count models, the EM Algorithm proposed by \cite{McLachlan1988fitting} will be utilized.  For truncated data sets described in \eqref{eqn:assumption}, fitting the model using EM algorithm is not straightforward as when all data points are available.  In general, the M-step of this algorithm does not have a closed form unless the complete data vector is extended to include indicator variables denoting the membership of data points with respect to the components of the mixture.
\\\\
\noindent If the null distribution is assumed to be ZIGP, then the log likelihood $\ell(\eta,\lambda,\theta \mid \boldsymbol{x}_N)$ of the entire data vector is

\begin{equation}\label{eqn:loglh}
\displaystyle \sum \limits_{j = 0}^{C} {n_j \log f_0 (j; \Theta)} + \displaystyle \sum \limits_{j = C + 1}^{K} {n_j \log f(j; \cdot)}
\end{equation}
\\\\
Suppose the sample space of $X$, denoted by $\mathcal{X}$, is partitioned into $K + 1$ mutually exclusive subsets $\mathcal{X}_j = \{j\}$, $j \in \cal A$, where independent observations are made on $X$.  After choosing a suitable value for $C$, the null sample $\boldsymbol{x}_n = (\boldsymbol{x}_0, \boldsymbol{x}_1, \ldots, \boldsymbol{x}_C)$ and the corresponding vector of mutation counts $\boldsymbol{y}_n = (n_0, n_1, \ldots, n_C)'$ are available for the estimation of the parameters of $f_0$.  However, the problem that arises is that the number of observations $n_j$ falling in $\mathcal{X}_j \hspace{2mm}, j > C$ are not available for the subsequent estimation of the parameters of $f_0$.\\
\\
\noindent For the $n$ observations in $\boldsymbol{x}_n$, it is assumed that $\boldsymbol{y}_n = (n_0, n_1, \ldots, n_C)'$ has a Multinomial distribution consisting of $n$ draws on $C + 1$ categories with probabilities $p_j$
\begin{equation}\label{eqn:multprob}
p_j = \displaystyle \frac {f_0 (j; \Theta)}{\displaystyle \sum \limits_{j = 0}^{C} f_0 (j; \Theta)}
\end{equation}
where $\Theta = (\eta, \lambda, \theta), \displaystyle \sum \limits_{j = 0}^{C} p_j = 1$ and $\displaystyle \sum \limits_{j = 0}^{C} {n_j} = n$.
\noindent This gives the likelihood function
\begin{eqnarray}\label{eqn:multlh}
L_0(\Theta; \boldsymbol{y}_n) = \displaystyle \frac{n!}{n_0! n_1! \ldots n_C!} \displaystyle \prod_{j = 0}^{C} {p_j^{n_j}}
\end{eqnarray}
From \eqref{eqn:multlh}, we can solve the likelihood equation $\partial L_0(\Theta ; \boldsymbol{y}_n)/\partial\Theta = \boldsymbol{0}$
within the EM framework following the work of \cite{Dempster1977maximum}.  The EM machinery is invoked by defining $\boldsymbol{w}_N = (\boldsymbol{y}'_n, \boldsymbol{y}'_{N - n})'$ as the complete-data vector where $\boldsymbol{y}_{N - n} = (n_{C + 1}, n_{C + 2}, \ldots n_K)'$.  Then, instead of looking at the log likelihood for $\boldsymbol{y}_n$, we consider the log likelihood function of the complete data, $\ell (\eta, \lambda, \theta \mid \boldsymbol{w}_N)$.  In order to find the estimates, it is important to note that each entry of $\boldsymbol{y}_{N - n}$ is a realization of a hidden random variable.  However, since these realizations do not exist in reality, we have to consider each entry of $\boldsymbol{y}_{N - n}$ as a random variable itself.\\
\\
Furthermore, \cite{McLachlan1988fitting} proposed an extension of the complete-data vector $\boldsymbol{w}_N$ for mixture densities to include the zero-one indicator variables
$$\boldsymbol{z}_{jk} = (z_{0jk}, z_{1jk})' \hspace{3mm} j = 0, 1,\ldots, K; k = 1, 2\ldots, n_j$$
where $z_{0jk} + z_{1jk} = 1$ and given the number of mutations $j$, $\boldsymbol{z}_{jk}$ are conditionally independent.  Conditional on the value of $j$, the probability of membership to a component can be computed using Bayes' Theorem as
\begin{eqnarray*}
\tau_{0j} (\Theta) = P(z_{0jk} = 1 \mid j) = \displaystyle \frac{\eta I_{\{0\}}(j)}{f_0 (j)}
\end{eqnarray*}
and $\tau_{1j} (\Theta) = P(z_{1jk} = 1 \mid j) = 1 - P(z_{0jk} = 1 \mid j)$.  The indicator function $I_{\mathcal{S}}(j)$ is equal to 1 if $j \in \mathcal{S}$ and 0 otherwise.
\newpage
\noindent
Using these indicator variables in the complete-data specification, the log likelihood becomes
\begin{eqnarray}\label{eqn:llh}
\sum \limits_{j = 0}^{K} \sum \limits_{k=1}^{n_j} z_{0jk}\log \eta I_{\{0\}}(j) +   \sum \limits_{j = 0}^{K} \sum \limits_{k=1}^{n_j} z_{1jk} \log \left[(1 - \eta)g(j)\right]
\end{eqnarray}
\\\\
The details of the EM Algorithm are provided in the Appendix.  Moreover, it is straightforward to estimate $f(j)$ by using relative frequency given by
\begin{eqnarray}\label{eqn:fest}
\hat{f}(j) = \displaystyle \frac{n_j}{n_0 + n_1 + \ldots + n_K}
\end{eqnarray}
\\
Using the assumption on $f_0$, for $j \hspace{1mm}\leq \hspace{1mm} C$, $f(j)$ from \eqref{eqn:f} reduces to $\pi_0 f_0 (j)$.  Hence,
\begin{eqnarray*}
	\displaystyle \sum \limits_{j = 0}^{C} \pi_0 f_0 (j) = \displaystyle \sum \limits_{j = 0}^{C} f(j)
\end{eqnarray*}
To estimate $\pi_0$, we need to calculate
\begin{eqnarray}
\hat{\pi}_0 = \displaystyle \frac{\displaystyle \sum \limits_{j = 0}^{C} {\hat{f}(j)}}{\displaystyle \sum \limits_{j = 0}^{C} {\hat{f}_0(j)}}
\end{eqnarray}
using \eqref{eqn:fest} and the estimate of $f_0$ after plugging in $\hat \Theta$ resulting from EM algorithm.
Finally, the estimate of $\pi_0$ is $min(1, \hat{\pi}_0)$.

\vspace{5mm}
\subsection{Choice of the Cut-off $C$}
\vspace{3mm}
In our model, we assume that we can identify a cut-off $C$, wherein bins with number of mutations greater than $C$ contain more mutations than what would be expected in the null model.  The choice of the cut-off $C$ is of paramount importance since the estimation of $f_0$ and $\pi_0$ depend on $C$.  It is more realistic to assume that $C$ is unknown, so such a predetermined $C$ may affect the result of local FDR procedure seriously.
\\\\
In particular, if $C$ is predetermined and is  chosen to be larger than the true value, the null distribution is estimated based on observations from alternative hypothesis as well as null hypothesis, so
the estimated null distribution is contaminated by the alternative distribution. This will cause insensitivity of local FDR procedure in detecting the alternative hypothesis.
On the other hand,  if $C$ is chosen to be smaller, then the null distribution is estimated only based on small values, so the estimation of the null distribution especially at the tail part is less reliable. Empirically, the FDR procedure yields liberal results in that
there are too many rejections resulting in failure in controlling a given level of FDR.
\newpage
\noindent
For the normal distribution as a null distribution,
\cite{Efron2007doing} proposed the maximum likelihood estimation from
likelihood based on observations in a given predetermined interval around zero.
\cite{Park2011estimation} considered a mixture of normal distributions for the null distribution and
proposed two approaches to select intervals around the mode to estimate the parameters in the mixture model using the EM algorithm.  One of the proposed methods is based on the idea of goodness of fit to the parametric model of the null distribution.
As the interval increases in length, it finally includes more and more alternative values resulting in
deviation from the null distribution.
\\\\
The estimation of the cut-off $C$ has been also formulated in the context of change-point analysis.  \cite{Sheetlin2011objective} offer an objective-change point method that can replace the subjective approaches performed by eye-balling the data.  The proposed method resembles the change-point regression and robust regression but it is tailored to estimate the change point from a transient to an asymptotic regime.  Given a tuning parameter $c$ and a criterion function $\rho$, depending on $\beta$, the estimator for the change point $k^{\star}$ is defined as
\begin{eqnarray}
k^{\star} = \argmin_{k = 0, 1, ..., n} \left( \min_{\beta}\displaystyle \sum \limits_{i = k + 1}^{n} {(\rho(e_i) - c)}\right)
\label{eqn:spouge}
\end{eqnarray}
where $\rho(e_i)$ is the estimated least-squares normalized residual.
In \eqref{eqn:spouge}, there is a tuning parameter $c$ which should be given ahead.  The value of $c$ plays the role of penalty for adding terms $\rho(e_i)$ in \eqref{eqn:spouge}, so the predetermined value of $c$ affects $k^{\star}$ arbitrarily.  We see that our proposed estimation of $C$ is related to the form \eqref{eqn:spouge}.
\\\\
We introduce our proposed estimation procedure of $C$.  Let us define the index sets
\begin{eqnarray}
{\cal A}= \{ j :  j \geq 0, n_j>0 \},  ~~~~~ {\cal A}(C) =  \{j : 0\leq j \leq C,  n_j>0, f_1(j)=0 \}.
\end{eqnarray}
Note that   $ {\cal A}(C_1) \subset {\cal A}(C_2)$   for $C_1 < C_2$   and    $f(j) =  \pi_0 f_0(j)$ when $j \in {\cal A}(C)$.
We adopt the idea of sequential testing to detect the change point in which
the observations are generated from the mixture distribution $f$.
More specifically, suppose we observed  $(0,n_0), (1, n_2), \ldots, (K, n_K)$ sequentially from
$f_0(0), f_0(1), \ldots, f_0(C), f(C+1),\ldots, f(K)$ where  distribution is changed from $f_0$  to $f$ at $C+1$.
\\\\
Our goal is to detect the change point $C$ based on assuming that we observe  $0,1,2,\ldots, K$ sequentially.
For a given $\nu$, we define  $S_{\nu}(\Theta)$ as
\begin{eqnarray}
S_{\nu} (\Theta ,f) =  \sum_{j\leq \nu} n_j \log f_0(j) + \sum_{j\geq \nu +1} n_j \log f(j) = \sum_{j\leq \nu} n_j \log \frac{f_0(j)}{f(j)} + \sum_{j\leq K} n_j \log f(j).
\end{eqnarray}
Maximizing $S_{\nu}(\Theta,f)$ is equivalent to the CUSUM(cumulative sum)  $\displaystyle \sum \limits_{j \leq \nu }  n_j \log \displaystyle \frac{f_0(j)}{f(j)}$.
Since the parameters $\Theta$ is estimated from EM algorithm and $\hat f(j)= \displaystyle \frac{n_j}{M}$,
our procedure is
\begin{eqnarray}
\hat C = {\rm argmax}_{\nu = 1, 2, \ldots, K} S_{\nu} (\hat \Theta_{\nu})
\end{eqnarray}
where $\hat \Theta_{\nu}$ is the estimator from the EM algorithm in the previous section with the value of $C$ set to $\nu$.  One may consider the full likelihood of all observations and find out some connection between $S_\nu$ and the full likelihood presented as follows.
\newpage
\noindent
The likelihood function of $(0, n_0), \ldots, (K, n_K)$ for a given ${\cal A}(\nu)$ is
\begin{eqnarray}
{\rm likelihood} \equiv  L(\Theta^\star, f) =\prod_{j \leq K} f(j)^{n_j} = \prod_{j \leq \nu}
(\pi_0 f_0(j))^{n_j} \prod_{j  \geq  \nu+1} f(j)^{n_j}
\end{eqnarray}
where  $\pi_0f_0$ depends on $\Theta^\star = (\pi_0, \eta, \theta, \lambda) = \{\pi_0\} \cup \Theta$.  The log likelihood   is also
\begin{eqnarray}
\log L(\Theta^\star, f)
= \ell_{\nu} (\Theta^\star,f) \equiv
\sum_{j \leq \nu} n_j \log (\pi_0 f_0(j)) + \sum_{j \geq \nu+1} n_j \log f(j)
\end{eqnarray}
since $f(j) = \pi_0 f_0(j)$ for $j \in {\cal A}(\nu)$. This leads to
\begin{eqnarray}
\ell_{\nu}  (\Theta^\star, f) &\equiv&   \sum_{j\leq \nu} n_j \log \frac{\pi_0f_0(j)}{f(j)}  + \sum_{j\leq K} n_j \log f(j)
\end{eqnarray}
which is equivalent to
\begin{eqnarray}
S_{\nu} (\Theta^\star, f) &=& \ell_{\nu} (\Theta^\star, f) - N_{\nu} \log \pi_0 \\
&=&   \sum_{j\leq \nu} n_j ({\rm lr}_j (\Theta^\star, f)-  \log \pi_0)   + l_0
\end{eqnarray}
where $N_{\nu}=\sum \limits_{j\leq \nu} n_j$ and
${\rm lr}_j (\Theta^\star,f) = \displaystyle \frac{\pi_0 f_0(j)}{f(j)}$.  It can be also seen that the penalized likelihood has the form of
\eqref{eqn:spouge}
\begin{eqnarray}
\sum_{j \in {\cal A}(C)} n_j (-{\rm lr}_j (\Theta^\star, f)-c)
\end{eqnarray}
where $c=-\log \pi_0$.
We estimate  $C$ via
\begin{eqnarray}
\hat C_1 &=&   {\rm argmin}_{\nu = 1, 2, \ldots, K}  \left( - S_{\nu} ( \hat \Theta^\star_{\nu}, \hat f)  \right) =
{\rm argmin}_{\nu = 1, 2, \ldots, K}     \left(  - \ell_{\nu}(\hat \Theta^\star_{\nu}, \hat f) + N_{\nu}\log \hat  \pi_{0, \nu}  \right) \\
&=&  {\rm argmin}_{\nu=1,2,\ldots, K}    \sum_{j \in \mathcal{A}(C)}  n_j \left(\rho_j(\hat \Theta^\star_{\nu}, \hat f)- \hat c_{\nu} \right)
\label{eqn:C1selection}
\end{eqnarray}
where $\hat{\Theta}^\star_\nu =( \hat \pi_{0,\nu}, \hat \eta_{\nu},  \hat \theta_{\lambda}, \hat \lambda_{\nu})$ is obtained from the EM algorithm discussed in the previous section,  $\hat f(j) = n_j/N$,
$\rho_j ( \hat \Theta^\star_{\nu}, \hat f)  = -{\rm lr}_{j} (\hat \Theta^\star_{\nu}, \hat f) $  and $\hat{c}_{\nu}= -\log \hat \pi_{0, \nu}$.
\\\\
In \eqref{eqn:spouge}, $c$ is a predetermined value, however  we don't need to predetermine any parameter in \eqref{eqn:C1selection}.  The proposed criterion \eqref{eqn:C1selection} is related to the penalized model selection such as AIC and BIC.  When we use the information that $n=\sum \limits_{j\leq C} n_j$ observed values are generated from  $f_0$,
$\sum \limits_{j\leq \nu} n_j {\rm lr}_j  (\Theta^\star,f)$ is increasing in $\nu$, there is a compromise term $c=-\log \pi_0$ for each observation to compensate  adding additional terms.  There is a total of $N_{\nu}$ positions, so when we use the assumption $\nu=C$,  we consider  $N_{\nu}\log \pi_0$ penalty to the log likelihood function $\ell_{\nu}$.  Most of well known model selection criteria have similar forms where the penalty terms
are related to penalize the complexity of models.  In our context, the term  $-\log \pi_0$ gives penalty to using the information that $j$ for $j \leq \nu$  are generated from $f_0$.  For a small value of $\pi_0$, the corresponding penalty ($-\log \pi_0$)  is large since a large penalty should be given to a low chance of $f_0$.  On the other hand, if $\pi_0$ is close to 1, there becomes small risk from assuming observations are from the null hypothesis.
\\\\
\newpage
\noindent
For the second method, we consider the extension of the methodology proposed by \cite{Efron2007doing} which explicitly uses the zero assumption. This stipulates that the non-null density $f_1$ is supported outside some set $\{0, 1, \ldots, C\}$.
 Let $n$ be the number
of mutations which is at most $C$ and define the likelihood function for $\boldsymbol{x}_n$ as
\begin{eqnarray*}
	L (\hat \Theta^\star_{\nu} \mid \boldsymbol{x}_n) = \xi^n (1 - \xi)^{N - n} \displaystyle \prod_{j \leq \nu} \left(f_0 (j)\right)^{n_j}
\end{eqnarray*}
where $\xi = \hat{\pi}_0 \sum \limits_{j = 0}^{C} {\hat{f}_0 (j)}$.  The cut-off can be computed as
\begin{eqnarray}
\hat C_2 &=& {\rm argmin}_{\nu=1,2,\ldots, K}    \left(\log  L(\hat \Theta^\star_{\nu} \mid \boldsymbol{x}_n)\right)
\label{eqn:C2selection}
\end{eqnarray}

\vspace{5mm}
\subsection{Modification of local FDR by truncation}
\vspace{3mm}
In practice, if a given domain position has a large number of mutations, then these mutations are expected to be significant.  In many cases,  there are relatively few positions in a protein domain where large values of mutations can be observed.  This indicates that for large values of $j$, estimation of  $f$ based on relative frequency is not accurate due to the sparse data in the tail part.  Consequently, the estimated local FDR is not reliable since it depends on the estimator of $f$.\\
\\
Rather than testing significance based on inaccurate local FDRs from large mutation counts, we consider a screening process so that the number of mutations exceeding a certain value should be considered as significant mutations.
Such a critical value will be decided depending on the estimated null distribution.  When we have observations $a_i$ for $1\leq i \leq N$ generated from the null distribution, we are interested in figuring out $D_N$ such that
\begin{eqnarray}
P \left( \max_{1\leq i \leq N} a_i < D_N \right) \rightarrow 1
\label{eqn:max1}
\end{eqnarray}
as $N \rightarrow \infty$.  Once a sequence $D_N$ is identified,  $a_i (\geq D_N)$ is hardly observed under the null hypothesis, so the corresponding null hypothesis is rejected directly rather than making decision based on local FDR procedure.
There are many choices of $D_N$, but a smaller sequence of $D_N$ satisfying \eqref{eqn:max1}
is of our interest since any sequence $B_N$  satisfying $B_N > D_N$
also satisfies the property.
\\\\
When $a_i$  is observed from Generalized Poisson distribution, \cite{Klar2000bounds} showed that the tail probabilities satisfy the following inequality:
\begin{eqnarray}
P(a_i \geq D_N) < \left[1 - e^{1 - \theta}\left(\theta + \displaystyle \frac{\lambda}{D_N + 1}\right)\right]^{-1} \displaystyle \frac{\lambda (\lambda + \theta D_N)^{D_N - 1}}{(D_N)^{D_N + 1/2}}e^{-\lambda - (\theta - 1) D_N} \label{eqn:max2}
\end{eqnarray}
where $D_N \geq \displaystyle \frac{\lambda}{e^{\theta - 1} - \theta}, \hspace{2mm}\theta \in (0, 1), \hspace{2mm}\lambda \hspace{1mm}\textgreater\hspace{1mm} 0$.
\newpage
\noindent
Using \eqref{eqn:max2}, we can compute for
\begin{eqnarray*}
P \left( \max_{1\leq i \leq N} a_i \geq D_N \right) &=& 1 - \left[1 - P(a_i \geq D_N)\right]^N \label{eqn:max3}\\
&\leq& 1 - \left[1 - (\delta_{D_N})^{-1}\displaystyle \frac{\lambda (\lambda + \theta D_N)^{D_N - 1}}{(D_N)^{D_N + 1/2}}e^{-\lambda - (\theta - 1) D_N} \right]^N \label{eqn:max4}
\end{eqnarray*}
where $\delta_{D_N} = 1 - e^{1 - \theta}\left(\theta + \displaystyle \frac{\lambda}{D_N + 1}\right)$.  For \eqref{eqn:max1} to hold,
\begin{eqnarray}
\log N - \log \delta_{D_N} + \log \left(\displaystyle \frac{\lambda (\lambda + \theta D_N)^{D_N - 1}}{(D_N)^{D_N + 1/2}}e^{-\lambda - (\theta - 1) D_N}\right) \rightarrow -\infty \label{eqn:max5}
\end{eqnarray}
and \eqref{eqn:max5} can be simplified in terms of $N$ and $D_N$ as
\begin{eqnarray*}
{\cal G}_N&\equiv& \log N - 0.5\log D_N - \log (D_N + 1) + D_N \log \left(\theta + \displaystyle \frac{\lambda}{D_N}\right) - (\theta - 1) D_N \label{eqn:max6}
\end{eqnarray*}
leading to
\begin{eqnarray}
{\cal G}_N \asymp \log N + (\log \theta -\theta + 1) D_N. \label{eqn:max7}
\end{eqnarray}
To assure that ${\cal G}_N \rightarrow -\infty$,
we can take   $ D_N = \zeta \log N$ for some constant $\zeta$ satisfying
$$\zeta \hspace{2mm}\textgreater\hspace{2mm} \displaystyle \frac{1}{\theta - 1 - \log \theta}$$
Since $\log \theta \leq \theta - 1$, then $\zeta \hspace{1mm}\textgreater\hspace{1mm} 0, \hspace{2mm} \theta \in (0, 1)$ as desired. Hence, we take
\begin{eqnarray}
D_N = \biggr \lceil \max\left(\displaystyle \frac{\lambda}{e^{\theta - 1} - \theta}, \hspace{2mm} \displaystyle \frac{\log N}{\theta - 1 - \log \theta}\right) \biggr \rceil \label{eqn:gpdn}
\end{eqnarray}
where  $\lceil x \rceil$ is the smallest integer greater than or equal to $x (x>0)$.  Meanwhile, if $a_i$  is observed from Poisson distribution, \cite{Mitzenmacher2005probability} derived the bounds for the tail probabilities using the Chernoff bound argument:
\begin{eqnarray}
P(a_i \geq D_N) < \displaystyle \frac{e^{-\lambda}(e\lambda)^{D_N}}{(D_N)^{D_N}} \label{eqn:pois1}
\end{eqnarray}
where $0 < \lambda < D_N$.  Using the inequality in \eqref{eqn:pois1},
\begin{eqnarray*}
	P \left( \max_{1\leq i \leq N} a_i \geq D_N \right) &\leq& 1 - \left(1 - \displaystyle \frac{e^{-\lambda}(e\lambda)^{D_N}}{(D_N)^{D_N}}\right)^N \label{eqn:pois2}
\end{eqnarray*}
and in order to satisfy the condition in \eqref{eqn:max1}, $\mathcal{P}_N \rightarrow -\infty$ where
\begin{eqnarray*}
\mathcal{P}_N \asymp \log N - D_N \log D_N \label{eqn:pois3}
\end{eqnarray*}
Therefore, we take
\begin{eqnarray}
D_N = \lceil \max\left(\displaystyle \lambda, \hspace{2mm} \log N \right) \rceil \label{eqn:pdn}
\end{eqnarray}
When $a_i$ is observed from ZIGP, $D_N$ can be calculated exactly as shown in \eqref{eqn:gpdn} since the derivation will eventually yield the leading terms in \eqref{eqn:max7} which does not involve $\eta$. Similarly, if $a_i$ is observed from ZIP, $D_N$ can be computed using \eqref{eqn:pdn}.

\vspace{5mm}
\subsection{Two Stage Procedure}
\vspace{3mm}
\noindent The proposed method can be summarized into two stages:
\begin{enumerate}
\item Using the likelihood method specified in Section 3.3, identify the cut-off point $C$.
\item Suppose $\hat{\Theta} = (\hat{\eta}, \hat{\lambda}, \hat{\theta})$ are the parameter estimates at the chosen $C$.  Using $\hat{\Theta}$, compute $D_N$ based on the specified formula in Section 3.4.  By construction, we expect the value of $D_N$ to fall within the interval $C < D_N \leq K$. However, it is probable to observe values of $D_N$ outside this interval.  Under these scenarios, we consider the following:
\begin{enumerate}[(a)]
	\item If the calculated value of $D_N$ exceeds $K$, we take $D_N = K$.  This implies that there is no screening process performed.
	\item  If the calculated value of $D_N$ is below $C$, we take $D_N = C + 1$.  This indicates that all values above the chosen $C$ are automatically declared as significant mutations.
\end{enumerate}
To incorporate these conditions on the formulation of $D_N$, we can modify \eqref{eqn:gpdn} as
\begin{eqnarray}
D_N = \min\left(\biggr \lceil \max\left(\displaystyle \frac{\lambda}{e^{\theta - 1} - \theta}, \hspace{2mm} \displaystyle \frac{\log N}{\theta - 1 - \log \theta}, \hspace{2mm}C + 1\right) \biggr \rceil, \hspace{2mm}K \right) \label{eqn:mgpdn}
\end{eqnarray}
and \eqref{eqn:pdn} as
\begin{eqnarray}
D_N = \min\left(\lceil \max\left(\displaystyle \lambda, \hspace{2mm} \log N, \hspace{2mm} C + 1 \right) \rceil, \hspace{2mm}K \right) \label{eqn:mpdn}
\end{eqnarray}
For a given null distribution, we can calculate $D_N$ using \eqref{eqn:mgpdn} or \eqref{eqn:mpdn} correspondingly.  After determining the value of $D_N$, all values of $j \geq D_N$ are considered significant mutations.
\end{enumerate}
\vspace{5mm}
Using this two-stage procedure, we can identify the mutation counts which are falsely rejected.  In the simulated data set, we can specify the value of true $C$.
As discussed previously, all mutation counts below $C$ are assumed to follow the null distribution $f_0$.  Hence, any rejection for mutation counts $j \leq C$ are considered to be erroneous.

\vspace{5mm}
\section{Numerical Studies}
\subsection{Simulation Studies}
\vspace{3mm}
\noindent To gain insights regarding the robustness of the proposed procedures in the presence of model misspecification, we perform some simulation studies.  The comparison is based on four simulation boundaries: (1) method used in the choice of the cut-off $C$; (2) model for the estimation of $f_0$; (3) null distribution; and (4) non-null distribution used in data generation.  There are two methods considered for the choice of cut-off $C$ as discussed in the previous section.  The null distributions considered are Zero-Inflated Poisson (ZIP) and Zero-Inflated Generalized Poisson (ZIGP) distribution. Both distributions account for the excessive number of zeros which is a characteristic of the mutation count data. For the non-null distribution, Geometric($p=0.08$) and Binomial($n=250, p=0.20$) distribution are utilized.  These were chosen because it can characterize the pattern of the mutation count observed in the real data set.
\newpage
\noindent
The assessment of the performance is also based on the model used in the estimation of $f_0$ since it affects calculation of the local FDR. The four models compared are ZIGP, ZIP, Generalized Poisson and Poisson distribution.  This allows for the comparison of the number of falsely rejected hypotheses when the model for $f_0$ is specified correctly and when there is departure from the true model of $f_0$.\\
\\
A total of $L$ hypotheses tests were performed for independent random variables $n_j$ over 1000 replications. For each replication, the proportion of $n_j$ from the null distribution is set to be $\pi_0$ and the total number of positions $N$ is specified to be 1000. To calculate the False Discovery Rate, $\widehat{FDR}$, for the $k$th generated data, $k = 1, 2, \ldots, 1000$, we compute the false discovery proportion (FDP) which is defined by
$$FDP_k = \displaystyle \frac{V_{k}}{R_{k}} I(R_{k} > 0)$$
where $V_{k}$ and $R_{k}$ are the number of falsely rejected hypotheses (false discoveries) and the total number of rejected hypotheses in the $k$th generated data, respectively.  FDR is the expected value of the false discovery proportion and can be computed empirically as
$$\widehat{FDR} = \displaystyle \frac{1}{1000} \displaystyle \sum \limits_{k = 1}^{1000} \displaystyle \frac{V_{k}}{R_{k}} I(R_{k} > 0)$$
In our simulations, the decision rule is to reject the null $H_{0j}$ if $fdr(j) = \hat{\pi_0}\hat{f}_0(j)/\hat{f}(j) < \alpha$.  Throughout the simulations, we consider  $\alpha = 0.05$.  The True Positive Rate, $\widehat{TPR}$ is computed empirically as
$$\widehat{TPR} = \displaystyle \frac{1}{1000} \displaystyle \sum \limits_{k = 1}^{1000} \displaystyle \left(\frac{S_{k}}{S_{k} + T_{k}}\right)$$
where $S_{k}$ and $T_{k}$ are the number of correctly rejected hypotheses (true discoveries) and the number of falsely accepted hypotheses (false non-discoveries) in the $k$th generated data, respectively.  Three procedures are compared in terms of controlling $\widehat{FDR}$ and $\widehat{TPR}$, namely the one-stage local FDR procedure, the proposed two-stage procedure and Storey's procedure.
\\\\
\scriptsize
\begin{center}
	\begin{tabular}{ccc}
		\begin{tabular}{c}
			\includegraphics[width=0.3\linewidth]{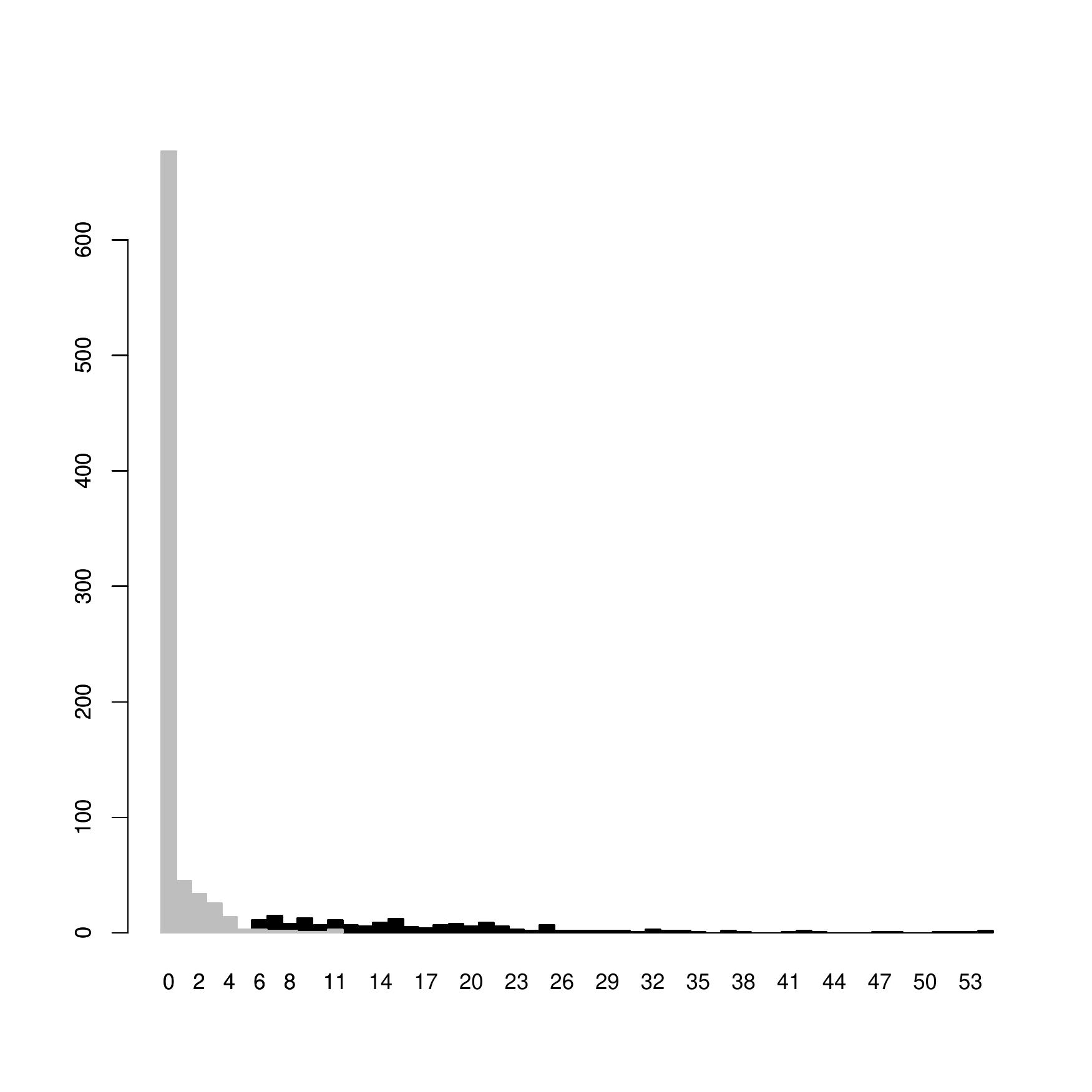}\\
			\textbf{ZIGP}$_1$\\
			ZIGP($\eta = 0.80, \lambda = 1.5, \theta=0.3$)\\
		\end{tabular}
		& \begin{tabular}{c}
			\includegraphics[width=0.3\linewidth]{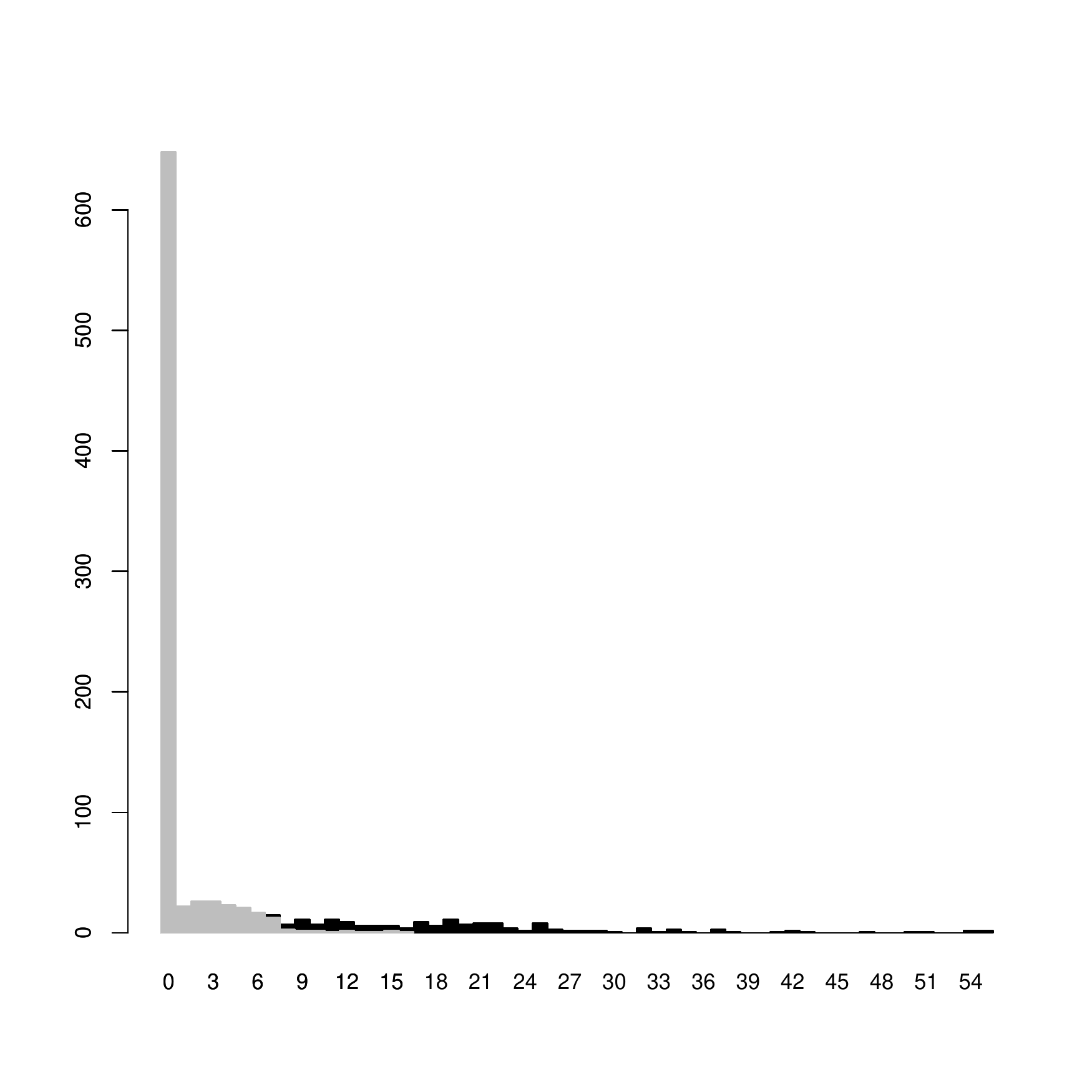}\\
			\textbf{ZIGP}$_2$\\
			ZIGP($\eta = 0.80, \lambda = 3, \theta = 0.3$)\\
		\end{tabular}
		& \begin{tabular}{c}
			\includegraphics[width=0.3\linewidth]{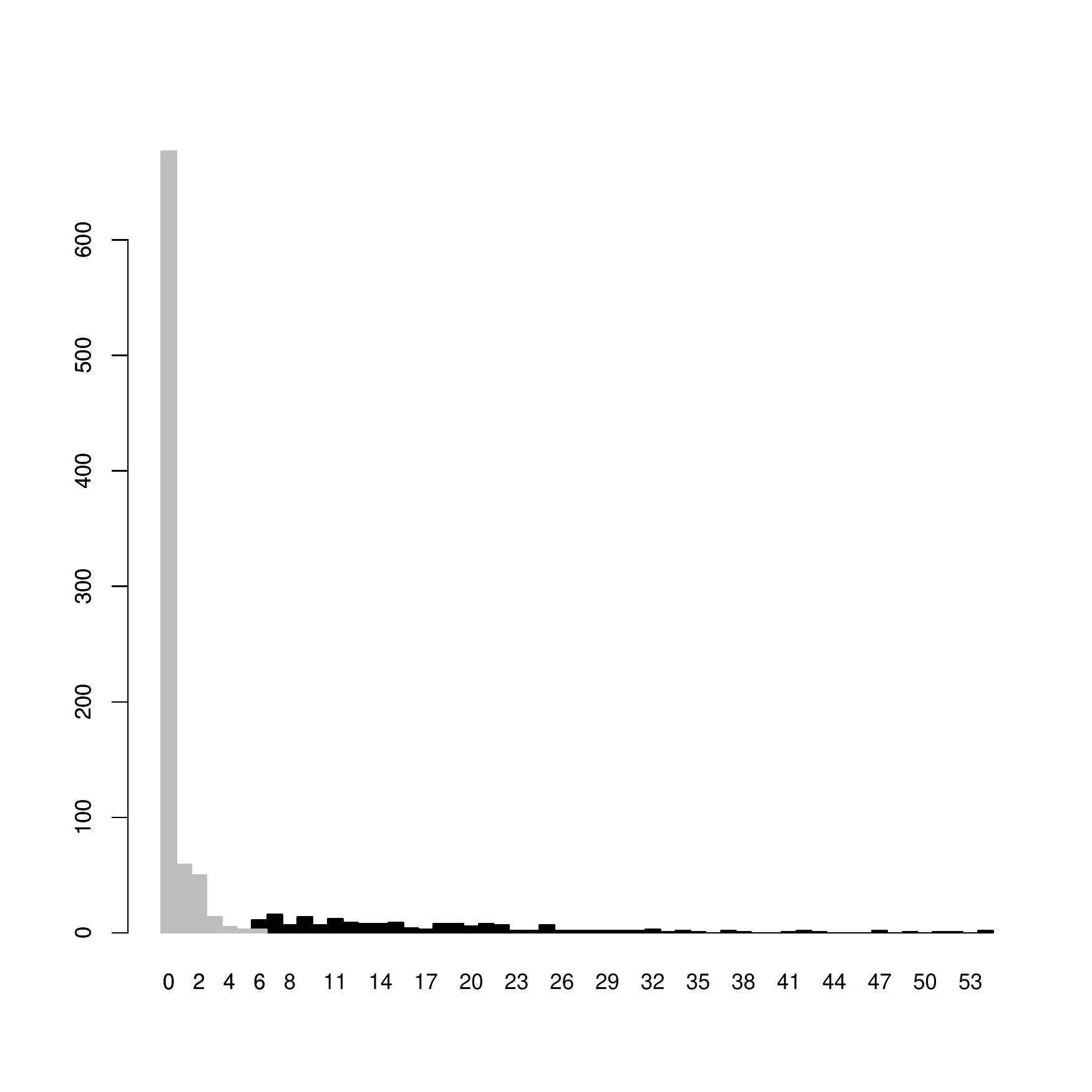}\\
			\textbf{ZIP}$_1$\\
			ZIP($\eta = 0.80, \lambda = 1.5$)\\
		\end{tabular}\\
	\end{tabular}\\
	\vspace{4mm}
	\textbf{Figure 1.} Histogram when the Non-null Distribution is Geometric($p=0.08$) and $\pi_0 = 0.80$.  ZIP$_1$ represents the well-separated case, ZIGP$_1$ is the moderately mixed case and ZIGP$_2$ is the heavily mixed case.
\end{center}
\newpage
\noindent
\normalsize
As displayed in Figure 1, the non-null distribution specified is Geometric($p=0.08$), $\pi_0 = 0.80$ and the fraction of zeros is 0.80.  The degree to which the null model is mixed with the non-null model is described using the three cases: ZIP$_1$, ZIGP$_1$ and ZIGP$_2$.  The corresponding numerical comparison is shown in Table 1.
\vspace{5mm}
\scriptsize
\begin{center}
	\textbf{Table 1.}  Numerical Comparison when the Non-null Distribution is Geometric($p=0.08$), $\pi_0 = 0.80$ and $\alpha = 0.05$.  The number in $(\cdot)$ represents the standard deviation.\\
	\vspace{3mm}
	\begin{tabular}{l|l|l|ccc|ccc|ccc}
		\hline
		\hline
		& & & \multicolumn{3}{c|}{Two-Stage Procedure} & \multicolumn{3}{c|}{One-Stage Procedure} &
		\multicolumn{3}{c}{Storey's FDR} \\
		\hline
		True & Choice & Model & & & &  & & &\\
		$f_0$ & of $C$& for $f_0$ & $R$ & $\widehat{FDR}$ & $\widehat{TPR}$ & $R$ & $\widehat{FDR}$ & $\widehat{TPR}$ & $R$ & $\widehat{FDR}$ & $\widehat{TPR}$ \\
		\hline
		&& &&& &&& &&\\
		ZIGP$_1$ & 	$C_1$ &ZIGP & 200.86 & 0.04422 & 0.95935 & 186.06 & 0.02992 & 0.90634 & 175.45 & 0.02185 & 0.85823 \\
		&&& (21.34) & (0.0196) & (0.0749) & (20.13) & (0.0150) & (0.0675) & (17.41) & (0.0123) & (0.0642) \\
		&&ZIP & 207.58 & 0.05193 & 0.98367 & 207.34 & 0.05176 & 0.98297 & 197.86 & 0.04102 & 0.94826 \\
		&&& (15.92) & (0.0217) & (0.0295) & (16.31) & (0.0219) & (0.0311) & (17.64) & (0.0203) & (0.0478) \\
		&&GP & 1.05 & 0.00000 & 0.00525 & 0.00 & 0.0000 & 0.00445 & 0.00 & 0.0000 & 0.00000 \\
		&&& (0.22) & (0.0000) & (0.0011) & (0.00) & (0.0000) & (0.0021) & (0.00) & (0.0000) & (0.0000) \\
		&&P & 278.88 & 0.28142 & 1.00000 & 278.88 & 0.28142 & 1.00000 & 253.81 & 0.21228 & 1.00000 \\
		&&& (19.55) & (0.0453) & (0.0000) & (19.55) & (0.0453) & (0.0000) & (14.06) & (0.0266) & (0.0000) \\
		&&& &&& &&& &&\\
		&$C_2$& 	ZIGP & 194.94 & 0.04008 & 0.93477 & 180.79 & 0.02717 & 0.88500 & 170.98 & 0.02003 & 0.83791 \\
		&&& (28.90) & (0.01999) & (0.11895) & (26.94) & (0.01525) & (0.10929) & (23.51) & (0.01244) & (0.10080) \\
		&&ZIP & 206.71 & 0.05095 & 0.98051 & 205.39 & 0.05010 & 0.97600 & 196.09 & 0.03983 & 0.94080 \\
		&&& (16.21) & (0.02195) & (0.03128) & (18.18) & (0.02288) & (0.04315) & (19.01) & (0.02086) & (0.05724) \\
		&&GP & 1.05 & 0.00000 & 0.00525 & 0.00 & 0.00000 & 0.00437 & 0.00 & 0.00000 & 0.00000 \\
		&&& (0.22) & (0.00000) & (0.00112) & (0.00) & (0.00000) & (0.00214) & (0.00) & (0.00000) & (0.00000) \\
		&&P & 278.88 & 0.28142 & 1.00000 & 278.88 & 0.28142 & 1.00000 & 253.81 & 0.21228 & 1.00000 \\
		&&& (19.55) & (0.04527) & (0.00000) & (19.55) & (0.04527) & (0.00000) & (14.06) & (0.02655) & (0.00000) \\ 		
		&&& &&& &&& &&\\
		\hline
		&& &&& &&& &&\\
		ZIGP$_2$ &$C_1$&ZIGP& 118.39 & 0.04694 & 0.55160 & 117.97 & 0.04646 & 0.55160 & 115.19 & 0.03739 & 0.54635 \\
		&&& (75.89) & (0.0419) & (0.3461) & (75.64) & (0.0417) & (0.3461) & (64.76) & (0.0335) & (0.2992) \\
		&&ZIP & 226.68 & 0.15855 & 0.95215 & 211.15 & 0.13399 & 0.91567 & 191.62 & 0.10364 & 0.85484 \\
		&&& (19.80) & (0.0359) & (0.0327) & (31.92) & (0.0529) & (0.0708) & (26.76) & (0.0421) & (0.07267) \\
		&&GP & 1.05 & 0.00000 & 0.00528 & 0.00 & 0.0000 & 0.00519 & 0.00 & 0.0000 & 0.00000 \\
		&&& (0.23) & (0.0000) & (0.0012) & (0.00) & (0.0000) & (0.0013) & (0.00) & (0.0000) & (0.0000) \\
		&&P & 334.22 & 0.40181 & 1.00000 & 334.22 & 0.40181 & 1.00000 & 316.02 & 0.36681 & 1.00000 \\
		&&& (15.36) & (0.0269) & (0.0000) & (15.36) & (0.0269) & (0.0000) & (17.55) & (0.0335) & (0.00000) \\
		&&& &&& &&& &&\\
		&$C_2$&ZIGP& 99.80 & 0.03745 & 0.46790 & 99.45 & 0.03701 & 0.46784 & 99.06 & 0.02964 & 0.47280 \\
		&&& (78.95) & (0.04110) & (0.36298) & (78.65) & (0.04082) & (0.36290) & (67.29) & (0.03273) & (0.31362) \\
		&&ZIP & 226.45 & 0.15823 & 0.95158 & 204.80 & 0.12532 & 0.89378 & 187.17 & 0.09803 & 0.83841 \\
		&&& (19.66) & (0.03582) & (0.03253) & (38.10) & (0.05961) & (0.09831) & (30.94) & (0.04620) & (0.09147) \\
		&&GP & 1.05 & 0.00000 & 0.00528 & 0.00 & 0.00000 & 0.00518 & 0.00 & 0.00000 & 0.00000 \\
		&&& (0.23) & (0.00000) & (0.00119) & (0.00) & (0.00000) & (0.00132) & (0.00) & (0.00000) & (0.00000) \\
		&&P & 334.22 & 0.40181 & 1.00000 & 334.22 & 0.40181 & 1.00000 & 316.02 & 0.36681 & 1.00000 \\
		&&& (15.36) & (0.02692) & (0.00000) & (15.36) & (0.02692) & (0.00000) & (17.55) & (0.03349) & (0.00000) \\
		&&& &&& &&& &&\\
		\hline
		&&& &&& &&& &&\\
		ZIP$_1$ &$C_1$& 	ZIGP & 200.85 & 0.00453 & 1.00000 & 187.43 & 0.00095 & 0.93620 & 182.76 & 0.00060 & 0.91352 \\
		&&& (13.09) & (0.0068) & (0.0000) & (14.19) & (0.0023) & (0.0260) & (13.87) & (0.0018) & (0.0356) \\
		&&ZIP & 198.14 & 0.00293 & 0.98782 & 198.13 & 0.00293 & 0.98782 & 191.24 & 0.00164 & 0.95467 \\
		&&& (14.47) & (0.0040) & (0.0243) & (14.48) & (0.0040) & (0.0243) & (15.19) & (0.0032) & (0.0371) \\
		&&GP & 1.04 & 0.00000 & 0.00525 & 0.00 & 0.0000 & 0.00321 & 0.00 & 0.0000 & 0.0000 \\
		&&& (0.22) & (0.0000) & (0.0011) & (0.00) & (0.0000) & (0.0026) & (0.00) & (0.0000) & (0.0000) \\
		&&P & 249.50 & 0.19300 & 1.00000 & 249.50 & 0.19300 & 1.0000 & 230.52 & 0.1327 & 1.0000 \\
		&&& (24.88) & (0.0731) & (0.0000) & (24.88) & (0.0731) & (0.0000) & (13.58) & (0.0232) & (0.0000) \\
		&&& &&& &&& &&\\
		&$C_2$ & 	ZIGP & 199.39 & 0.00453 & 0.99268 & 182.37 & 0.00061 & 0.91191 & 174.91 & 0.00029 & 0.87453 \\
		&&& (13.90) & (0.00632) & (0.02290) & (14.89) & (0.00182) & (0.03567) & (13.70) & (0.00123) & (0.03660) \\
		&&ZIP & 195.07 & 0.00234 & 0.97312 & 194.28 & 0.00229 & 0.96964 & 188.18 & 0.00146 & 0.93966 \\
		&&& (15.06) & (0.00370) & (0.03339) & (15.81) & (0.00370) & (0.03918) & (16.14) & (0.00305) & (0.04872) \\
		&&GP & 1.04 & 0.00000 & 0.00525 & 0.00 & 0.00000 & 0.00312 & 0.00 & 0.00000 & 0.00000 \\
		&&& (0.22) & (0.00000) & (0.00113) & (0.00) & (0.00000) & (0.00262) & (0.00) & (0.00000) & (0.00000) \\
		&&P & 249.50 & 0.19300 & 1.00000 & 249.50 & 0.19300 & 1.00000 & 230.52 & 0.13270 & 1.00000 \\
		&&& (24.88) & (0.07306) & (0.00000) & (24.88) & (0.07306) & (0.00000) & (13.58) & (0.02318) & (0.00000) \\
		&&& &&& &&& &&\\
		\hline
		\hline
	\end{tabular}
\end{center}
\newpage
\noindent
\normalsize
Overall, there are more rejections using $C_1$ as a cut-off compared to $C_2$.  This suggests that the extension of Efron's method is conservative and would miss significant positions.  Also, even if using $C_1$ yields more rejections, it sill controls the value of FDR indicating the superiority of $C_1$ as a cut-off method.
\\\\
The difference between $C_1$ and $C_2$ is further highlighted for ZIGP$_2$, where the true null distribution is heavily mixed with the non-null distribution and overdispersion is also present.  When the model used for the estimation of $f_0$ is ZIGP, the value of $\widehat{TPR}$ is relatively higher using $C_1$, while keeping the $\widehat{FDR}$ controlled.
\\\\
The results for the three null models can also be compared.  Since null and non-null distribution is moderately mixed for ZIGP$_1$, the resulting $\widehat{TPR}$ for all three procedures is substantially higher than the $\widehat{TPR}$ for ZIGP$_2$, regardless of the model used for the estimation of $f_0$.  Given that $\widehat{FDR}$ is controlled in all procedures if the model for $f_0$ is ZIGP, the Two-Stage procedure yields the highest $\widehat{TPR}$ compared to the One-Stage local FDR and Storey's procedure.  This suggests that the proposed procedure is better than the other existing procedures.
\\\\
Meanwhile, ZIGP$_1$ is allowed to vary from ZIP$_1$ in terms of the overdispersion parameter $\theta$.  Due to the ``well-separation" if the true null is ZIP$_1$, then the $\widehat{TPR}$ for ZIP$_1$ is slightly higher than the $\widehat{TPR}$ for ZIGP$_1$.  Moreover, the $\widehat{FDR}$ for all three procedures for ZIP$_1$ are noticeably lower than the $\widehat{FDR}$ for ZIGP$_1$.  This means that the number of rejections for ZIGP$_1$ and ZIP$_1$ are almost the same but there are more erroneous rejections for ZIGP$_1$.  This result can be explained by the presence of overdispersion in ZIGP$_1$ as compared to ZIP$_1$.
\\\\
Figure 2 presents the histograms when the non-null distribution specified is Binomial, the proportion of null cases is 0.80 and the fraction of zeros is 0.40.  Unlike the parametrization of the Geometric non-null distribution which appears to be skewed to the right, this non-null distribution exhibits near symmetry.  Similar to the previous set of results, the true null distribution is allowed to vary in terms of $\lambda$ and $\theta$.  In terms of the mixing of the null and non-null distribution, ZIP$_2$ represents the well-separated case, ZIGP$_3$ is the moderately mixed case while ZIGP$_4$ can be described as the heavily mixed case.  The respective numerical comparison is shown in Table 2.

\scriptsize
\begin{center}
	\begin{tabular}{ccc}
		\begin{tabular}{c}
			\includegraphics[width=0.3\linewidth]{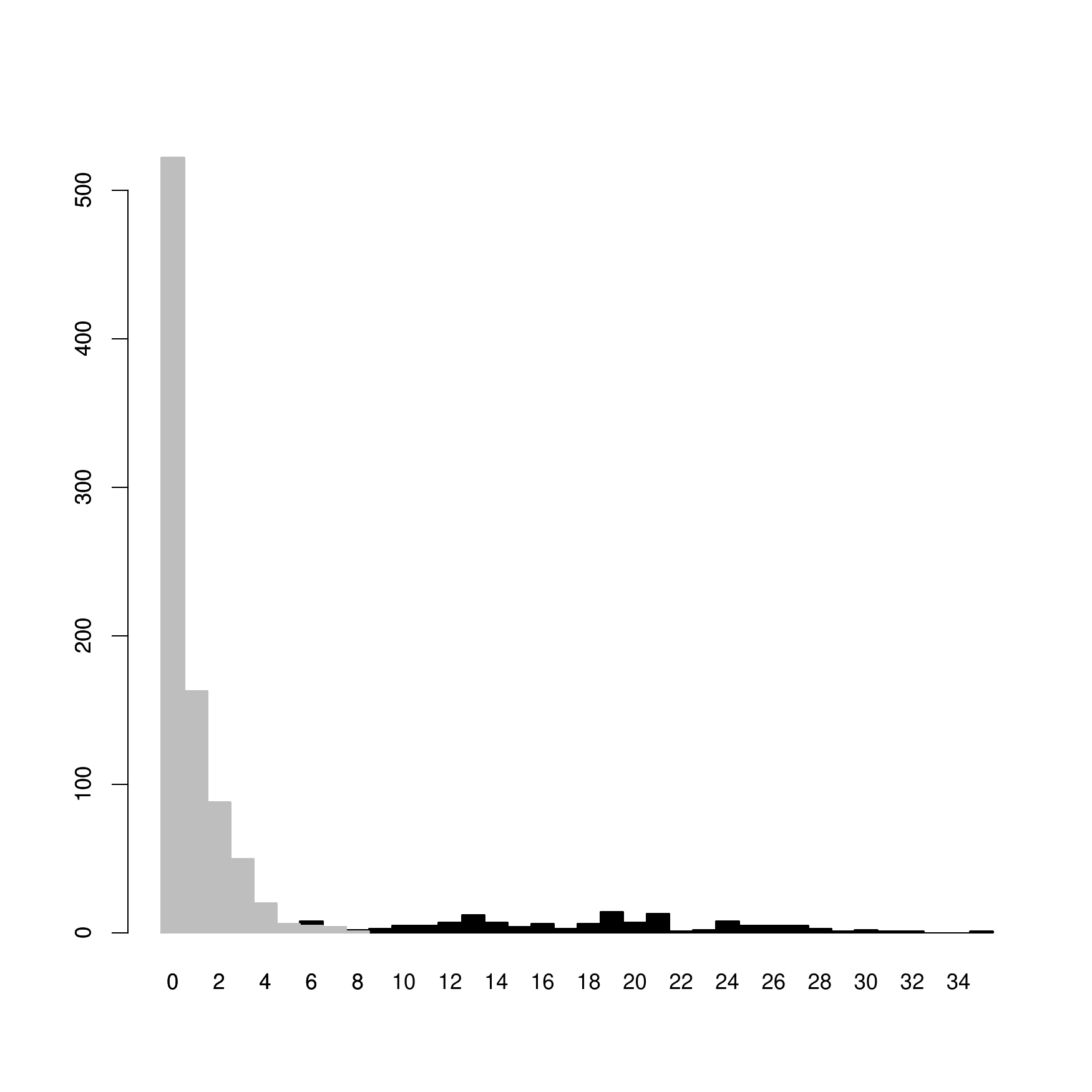}\\
			\textbf{ZIGP}$_3$\\
			ZIGP($\eta = 0.40, \lambda = 1, \theta=0.20$)\\
		\end{tabular}
		& \begin{tabular}{c}
			\includegraphics[width=0.3\linewidth]{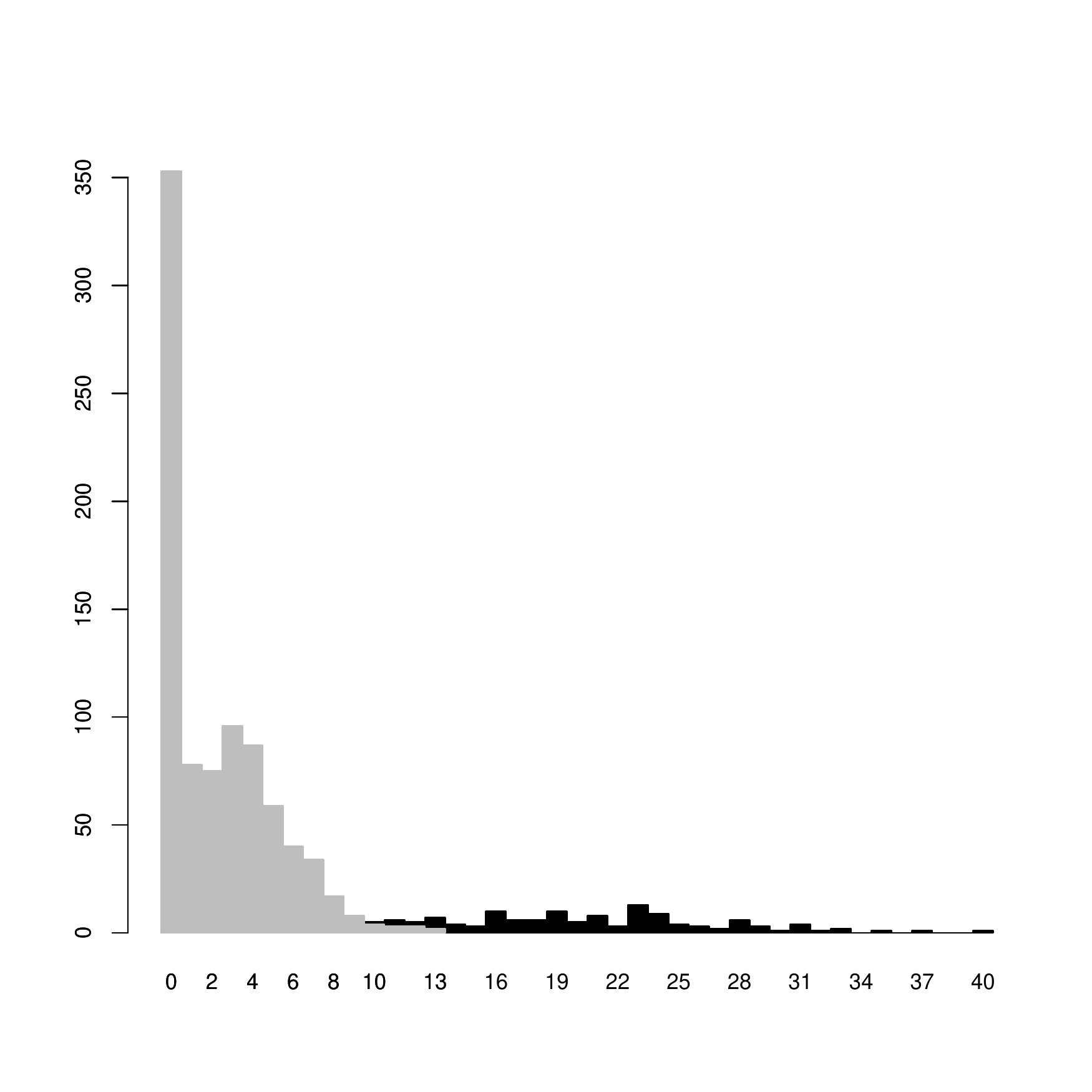}\\
			\textbf{ZIGP}$_4$\\
			ZIGP($\eta = 0.40, \lambda = 3, \theta=0.20$)\\
		\end{tabular}
		& \begin{tabular}{c}
			\includegraphics[width=0.3\linewidth]{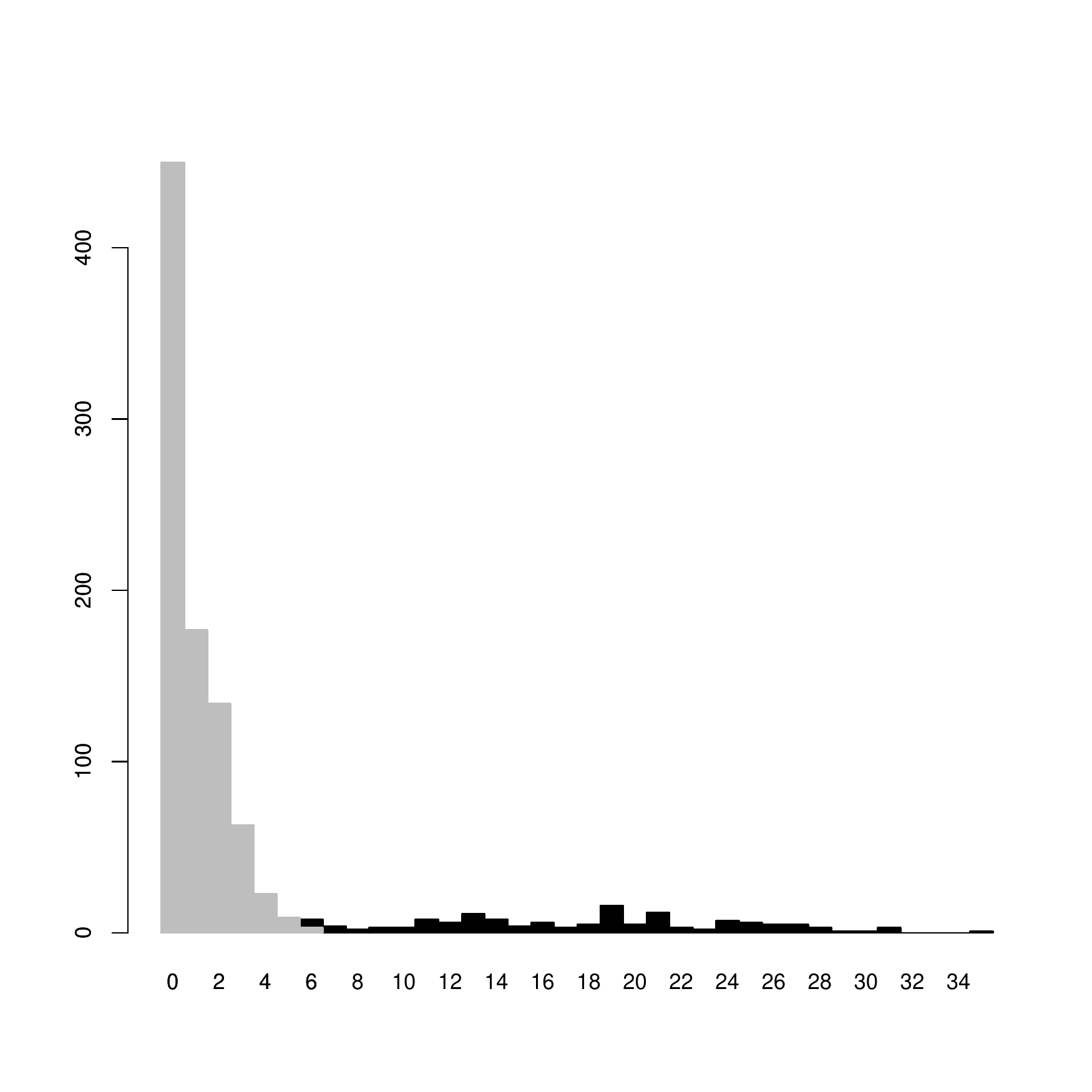}\\
			\textbf{ZIP}$_2$\\
			ZIP($\eta = 0.40, \lambda = 1.5$)\\
		\end{tabular}\\
	\end{tabular}\\
	\vspace{5mm}
	\textbf{Figure 2.} Histogram when the Non-null Distribution is Binomial($n=250, p=0.20$) and $\pi_0 = 0.80$.  ZIP$_2$ represents the well-separated case, ZIGP$_3$ is the moderately mixed case and ZIGP$_4$ is the heavily mixed case.
\end{center}
\vspace{5mm}
\noindent
\newpage
\scriptsize
\begin{center}
	\textbf{Table 2.}  Numerical Comparison when the Non-null Distribution is Binomial($n=250, p=0.20$), $\pi_0 = 0.80$ and $\alpha = 0.05$.  The number in $(\cdot)$ represents the standard deviation.\\
	\vspace{3mm}
	\begin{tabular}{l|l|l|ccc|ccc|ccc}
		\hline
		\hline
		& & &\multicolumn{3}{c|}{Two-Stage Procedure} & \multicolumn{3}{c|}{One-Stage Procedure} &
		\multicolumn{3}{c}{Storey's FDR} \\
		\hline
		True &Choice & Model & & & &  & & &\\
		$f_0$ & of $C$ & for $f_0$ & $R$ & $\widehat{FDR}$ & $\widehat{TPR}$ & $R$ & $\widehat{FDR}$ & $\widehat{TPR}$ & $R$ & $\widehat{FDR}$ & $\widehat{TPR}$ \\
		\hline
		&& &&& &&& &&\\
		ZIGP$_3$ &$C_1$& 	ZIGP & 209.45 & 0.04586 & 1.00000 & 186.45 & 0.00647 & 0.92738 & 190.90 & 0.01274 & 0.94354 \\
		&&& (13.76) & (0.02707) & (0.00000) & (12.64) & (0.00646) & (0.01928) & (12.81) & (0.00886) & (0.01817) \\
		&&ZIP & 203.71 & 0.02861 & 0.99039 & 203.57 & 0.02835 & 0.99039 & 203.38 & 0.02848 & 0.98897 \\
		&&& (14.31) & (0.01380) & (0.01957) & (14.45) & (0.01405) & (0.01957) & (14.35) & (0.01405) & (0.02258) \\
		&&GP & 68.03 & 0.00083 & 0.34534 & 65.57 & 0.00039 & 0.33245 & 72.13 & 0.00044 & 0.36634 \\
		&&& (80.14) & (0.00412) & (0.40543) & (78.36) & (0.00166) & (0.39491) & (81.72) & (0.00197) & (0.41397) \\
		&&P & 255.54 & 0.21280 & 1.00000 & 255.54 & 0.21280 & 1.00000 & 234.19 & 0.14702 & 1.00000 \\
		&&& (25.60) & (0.07095) & (0.00000) & (25.60) & (0.07095) & (0.00000) & (13.84) & (0.02419) & (0.00000) \\
		&&& &&& &&& &&\\
		&$C_2$ & 	ZIGP & 209.68 & 0.04703 & 1.00000 & 185.32 & 0.00550 & 0.92263 & 188.87 & 0.00990 & 0.93613 \\
		&&& (13.76) & (0.02436) & (0.00000) & (12.75) & (0.00619) & (0.02064) & (12.94) & (0.00809) & (0.02037) \\
		&&ZIP & 203.79 & 0.02875 & 0.99069 & 203.65 & 0.02849 & 0.99069 & 203.68 & 0.02888 & 0.99004 \\
		&&& (14.15) & (0.01382) & (0.01959) & (14.30) & (0.01407) & (0.01959) & (14.17) & (0.01406) & (0.02183) \\
		&&GP & 64.52 & 0.00082 & 0.32815 & 62.08 & 0.00041 & 0.31511 & 67.48 & 0.00055 & 0.34355 \\
		&&& (79.42) & (0.00412) & (0.40250) & (77.63) & (0.00170) & (0.39230) & (81.83) & (0.00219) & (0.41528) \\
		&&P & 255.54 & 0.21280 & 1.00000 & 255.54 & 0.21280 & 1.00000 & 234.19 & 0.14702 & 1.00000 \\
		&&& (25.60) & (0.07095) & (0.00000) & (25.60) & (0.07095) & (0.00000) & (13.84) & (0.02419) & (0.00000) \\
		&&& &&& &&& &&\\
		\hline
		&&& &&& &&& &&\\
		ZIGP$_4$ &$C_1$&ZIGP& 174.26 & 0.04521 & 0.82936 & 160.76 & 0.02237 & 0.82550 & 173.01 & 0.03546 & 0.83355 \\
		&&& (43.43) & (0.03435) & (0.19135) & (41.06) & (0.02015) & (0.19064) & (33.78) & (0.02505) & (0.14681) \\
		&&ZIP & 254.22 & 0.24384 & 0.96212 & 205.90 & 0.10126 & 0.93342 & 206.28 & 0.10221 & 0.92392 \\
		&&& (15.19) & (0.02873) & (0.01405) & (26.80) & (0.06698) & (0.02812) & (19.78) & (0.04791) & (0.02593) \\
		&&GP & 1.20 & 0.00000 & 0.00604 & 0.00 & 0.00000 & 0.00604 & 13.22 & 0.00000 & 0.06649 \\
		&&& (0.53) & (0.00000) & (0.00265) & (0.00) & (0.00000) & (0.00265) & (6.14) & (0.00000) & (0.03121) \\
		&&P & 538.18 & 0.62710 & 1.00000 & 538.18 & 0.62710 & 1.00000 & 476.05 & 0.57682 & 1.00000 \\
		&&& (41.90) & (0.03245) & (0.00000) & (41.90) & (0.03245) & (0.00000) & (47.22) & (0.04521) & (0.00000) \\
		&&& &&& &&& &&\\
		&$C_2$ &ZIGP& 98.53 & 0.02540 & 0.47057 & 89.99 & 0.01500 & 0.46826 & 115.13 & 0.02001 & 0.55907 \\
		&&& (89.67) & (0.03596) & (0.42354) & (84.25) & (0.02083) & (0.42133) & (69.14) & (0.02697) & (0.32719) \\
		&&ZIP & 254.22 & 0.24384 & 0.96212 & 204.28 & 0.10051 & 0.92299 & 205.19 & 0.10092 & 0.91887 \\
		&&& (15.19) & (0.02873) & (0.01405) & (30.40) & (0.07169) & (0.05145) & (22.79) & (0.05353) & (0.03732) \\
		&&GP & 1.20 & 0.00000 & 0.00604 & 0.00 & 0.00000 & 0.00604 & 8.34 & 0.00000 & 0.04179 \\
		&&& (0.53) & (0.00000) & (0.00265) & (0.00) & (0.00000) & (0.00265) & (3.35) & (0.00000) & (0.01653) \\
		&&P & 538.18 & 0.62710 & 1.00000 & 538.18 & 0.62710 & 1.00000 & 476.05 & 0.57682 & 1.00000 \\
		&&& (41.90) & (0.03245) & (0.00000) & (41.90) & (0.03245) & (0.00000) & (47.22) & (0.04521) & (0.00000) \\
		&&& &&& &&& &&\\
		\hline
		&&& &&& &&& &&\\
		ZIP$_2$ &$C_1$& 	ZIGP & 201.82 & 0.01046 & 0.99983 & 184.85 & 0.00028 & 0.92513 & 187.86 & 0.00110 & 0.93943 \\
		&&& (12.93) & (0.00823) & (0.00170) & (12.68) & (0.00126) & (0.02015) & (12.85) & (0.00270) & (0.01981) \\
		&&ZIP & 190.88 & 0.00261 & 0.95307 & 188.15 & 0.00156 & 0.95103 & 190.92 & 0.00266 & 0.95327 \\
		&&& (13.01) & (0.00456) & (0.01783) & (13.32) & (0.00421) & (0.01763) & (12.92) & (0.00449) & (0.01957) \\
		&&GP & 195.38 & 0.00689 & 0.97134 & 179.43 & 0.00015 & 0.89725 & 184.31 & 0.00067 & 0.92223 \\
		&&& (15.79) & (0.00790) & (0.04563) & (13.93) & (0.00090) & (0.04024) & (13.40) & (0.00221) & (0.03359) \\
		&&P & 244.13 & 0.17402 & 1.00000 & 244.13 & 0.17402 & 1.00000 & 222.66 & 0.10062 & 1.00000 \\
		&&& (28.99) & (0.07862) & (0.00000) & (28.99) & (0.07862) & (0.00000) & (17.54) & (0.04938) & (0.00000) \\
		&&& &&& &&& &&\\
		&$C_2$ & 	ZIGP & 201.55 & 0.01087 & 0.99807 & 181.79 & 0.00010 & 0.90989 & 186.02 & 0.00038 & 0.93100 \\
		&&& (13.13) & (0.00922) & (0.01027) & (12.63) & (0.00073) & (0.02130) & (12.50) & (0.00147) & (0.01900) \\
		&&ZIP & 190.93 & 0.00263 & 0.95328 & 187.68 & 0.00147 & 0.94782 & 190.62 & 0.00259 & 0.95181 \\
		&&& (13.01) & (0.00464) & (0.01796) & (13.40) & (0.00420) & (0.01967) & (13.11) & (0.00465) & (0.02136) \\
		&&GP & 195.19 & 0.00714 & 0.97024 & 178.59 & 0.00015 & 0.89298 & 183.63 & 0.00066 & 0.91884 \\
		&&& (18.21) & (0.00785) & (0.06445) & (16.38) & (0.00091) & (0.05861) & (15.85) & (0.00220) & (0.05344) \\
		&&P & 244.13 & 0.17402 & 1.00000 & 244.13 & 0.17402 & 1.00000 & 222.66 & 0.10062 & 1.00000 \\
		&&& (28.99) & (0.07862) & (0.00000) & (28.99) & (0.07862) & (0.00000) & (17.54) & (0.04938) & (0.00000) \\
		&&& &&& &&& &&\\
		\hline
		\hline
	\end{tabular}
\end{center}
\newpage
\noindent
\normalsize
\normalsize
The difference between $C_1$ and $C_2$ is apparent for ZIGP$_4$, where there is overdispersion and the true null distribution is heavily mixed with the non-null distribution.  If ZIGP is the model used for the estimation of $f_0$, the value of $\widehat{TPR}$ is substantially higher using $C_1$, while keeping the $\widehat{FDR}$ controlled.
\\\\
According to Table 2, the resulting $\widehat{TPR}$ for ZIGP$_3$ is substantially higher than the $\widehat{TPR}$ for ZIGP$_4$, regardless of the model used for the estimation of $f_0$ and the procedure employed. Given that $\widehat{FDR}$ is controlled in all procedures for ZIGP$_3$ if the model used for the estimation of $f_0$ is ZIGP, this suggests that the Two-Stage procedure is better than the One-Stage procedure and Storey's procedure.  However, for the scenario specified in ZIGP$_4$, the Storey's procedure is slightly better than the Two-Stage procedure if ZIGP is the model used for $f_0$.
\\\\
It can also be noted that for ZIP$_2$, the number of erroneous rejections is lesser if the model used for the estimation of $f_0$ is ZIP as compared to ZIGP.  However, given that $\widehat{FDR}$ is controlled by specifying either of the two models, using ZIGP leads to a higher $\widehat{TPR}$ than when the true model ZIP is specified. This result implies using ZIGP would yield satisfactory results even under model misspecification.
\\\\
As presented in Figure 3, the non-null distribution considered is also Binomial, fraction of zeros is still 0.40 but the proportion of null cases is reduced to 0.35.  Again, the true null distribution is allowed to vary in terms of $\lambda$ and $\theta$.  ZIP$_3$ represents the well-separated case, ZIGP$_5$ is the moderately mixed case while ZIGP$_{6}$ can be described as the overdispersed and heavily mixed case.  The respective numerical comparison is shown in Table 3.
\\\\
Based on the results shown in Table 3, using $C_1$ as a cut-off resulted to more rejections in the case of ZIGP$_5$ and ZIP$_3$.  However, contrary to the results from Table 1 and 2, there are more rejections using $C_2$ for ZIGP$_{6}$, where there is overdispersion and the true null distribution is heavily mixed with the non-null distribution.  If ZIGP is the model used for the estimation of $f_0$, the value of $\widehat{TPR}$ is substantially higher using Storey's procedure, while keeping the $\widehat{FDR}$ controlled.
\\\\
Furthermore, the resulting $\widehat{TPR}$ for ZIGP$_5$ is substantially higher than the $\widehat{TPR}$ for ZIGP$_{6}$, regardless of the model used for the estimation of $f_0$ and the procedure employed.
\vspace{5mm}
\scriptsize
\begin{center}
	\begin{tabular}{ccc}
		\begin{tabular}{c}
			\includegraphics[width=0.3\linewidth]{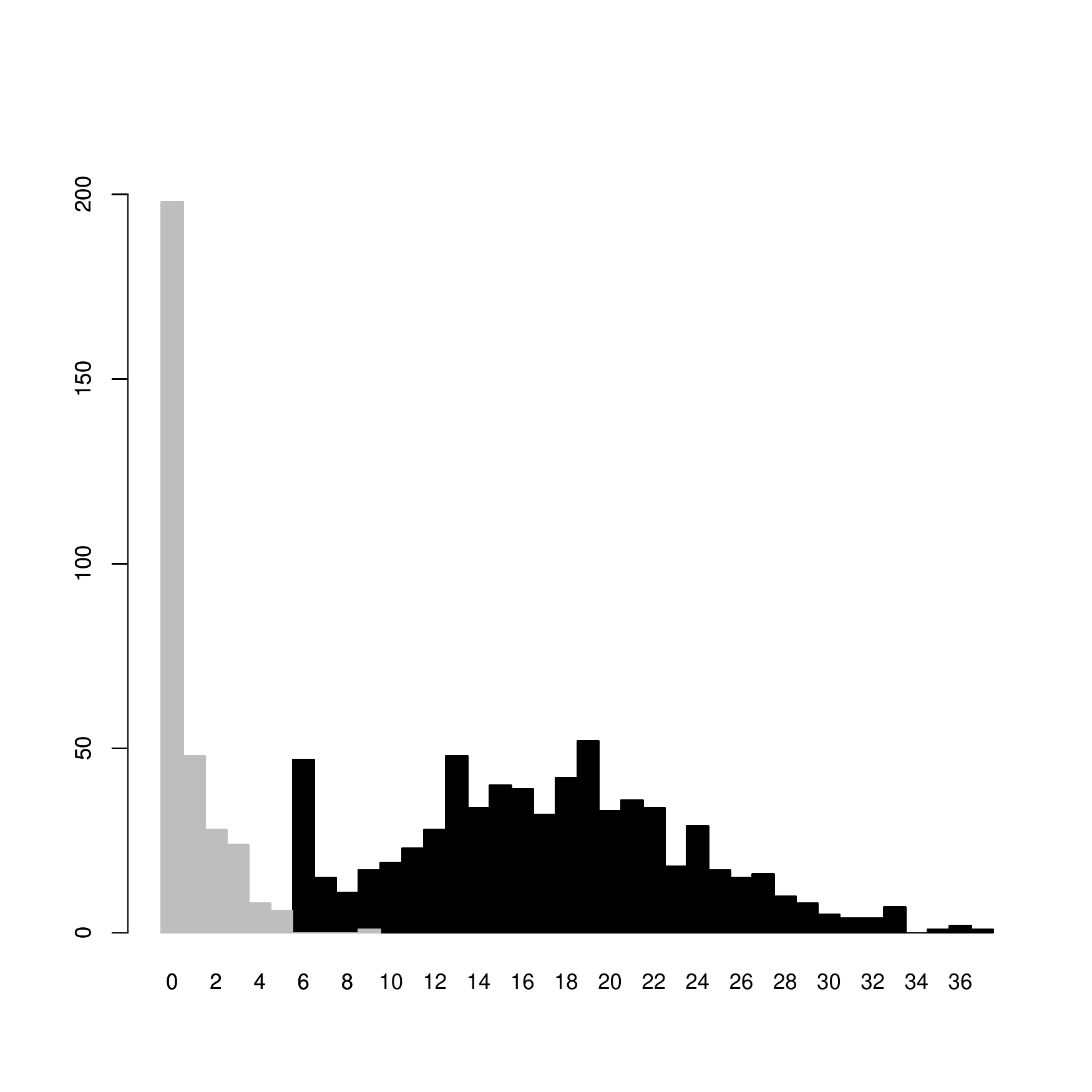}\\
			\textbf{ZIGP}$_{5}$\\
			ZIGP($\eta = 0.40, \lambda = 1, \theta=0.30$)\\
		\end{tabular}
		& \begin{tabular}{c}
			\includegraphics[width=0.3\linewidth]{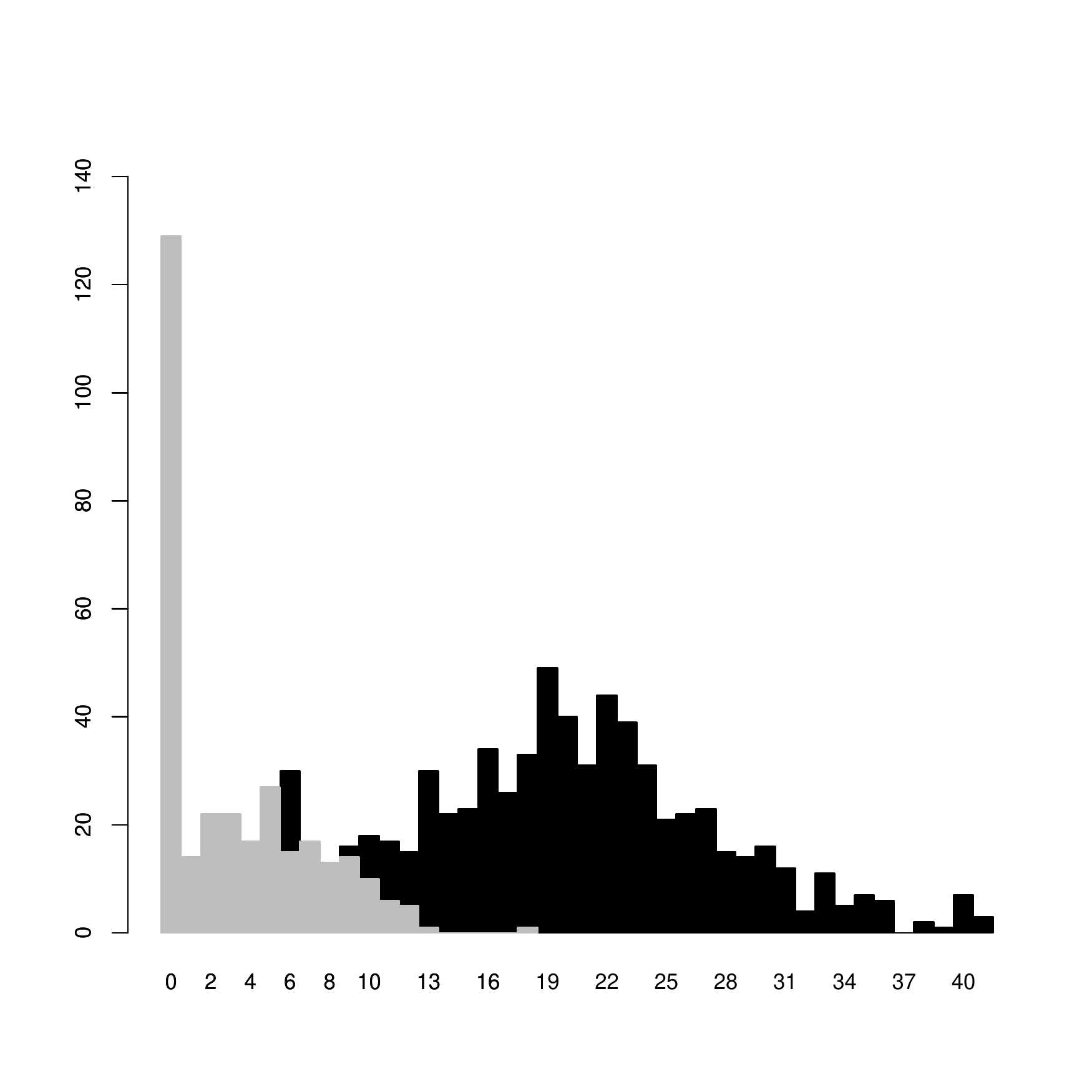}\\
			\textbf{ZIGP}$_{6}$\\
			ZIGP($\eta = 0.40, \lambda = 4, \theta=0.30$)\\
		\end{tabular}
		& \begin{tabular}{c}
			\includegraphics[width=0.3\linewidth]{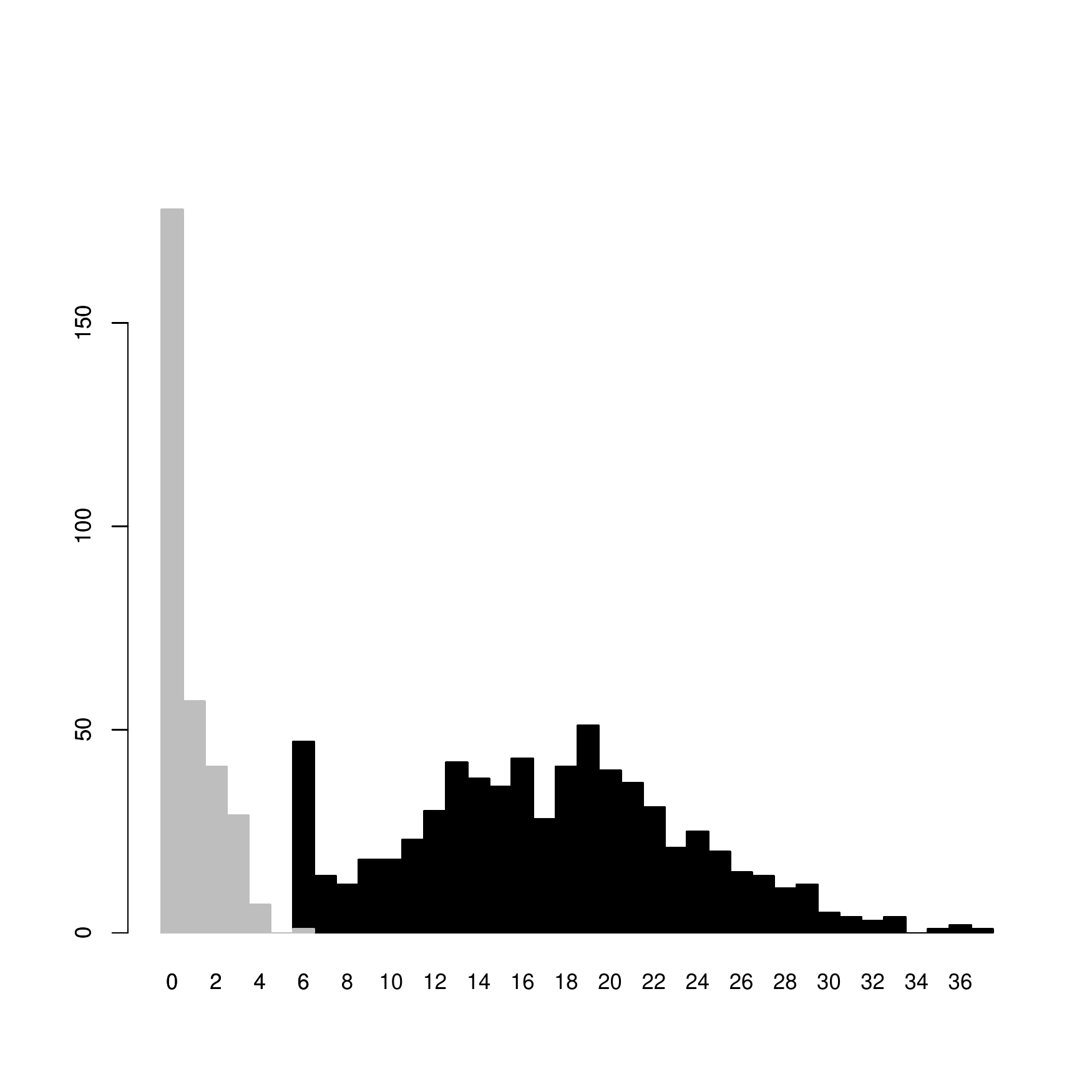}\\
			\textbf{ZIP}$_3$\\
			ZIP($\eta = 0.40, \lambda = 1.5$)\\
		\end{tabular}\\
	\end{tabular}\\
	\vspace{5mm}
	\textbf{Figure 3.} Histogram when the Non-null Distribution is Binomial($n=250, p=0.20$) and $\pi_0 = 0.35$.  ZIP$_3$ represents the well-separated case, ZIGP$_5$ is the moderately mixed case and ZIGP$_6$ is the heavily mixed case.
\end{center}
\newpage
\noindent
\begin{center}
	\textbf{Table 3.}  Numerical Comparison when the Non-null Distribution is Binomial($n=250, p=0.20$), $\pi_0 = 0.35$ and $\alpha = 0.05$.  The number in $(\cdot)$ represents the standard deviation.\\
	\vspace{3mm}
	\begin{tabular}{l|l|l|ccc|ccc|ccc}
		\hline
		\hline
		&&& \multicolumn{3}{c|}{Two-Stage Procedure} & \multicolumn{3}{c|}{One-Stage Procedure} &
		\multicolumn{3}{c}{Storey's FDR} \\
		\hline
		True &Choice & Model & & & &  & & &\\
		$f_0$ & of $C$ & for $f_0$ & $R$ & $\widehat{FDR}$ & $\widehat{TPR}$ & $R$ & $\widehat{FDR}$ & $\widehat{TPR}$ & $R$ & $\widehat{FDR}$ & $\widehat{TPR}$ \\
		\hline
		&& &&& &&& &&\\
		ZIGP$_{5}$ &$C_1$& ZIGP & 655.70 & 0.01105 & 0.99796 & 645.10 & 0.00886 & 0.98449 & 621.88 & 0.00620 & 0.95109 \\
		&&& (15.99) & (0.00543) & (0.01035) & (24.24) & (0.00450) & (0.02442) & (18.86) & (0.00355) & (0.01700) \\
		&&ZIP & 656.25 & 0.01040 & 0.99945 & 656.21 & 0.01039 & 0.99945 & 643.14 & 0.00894 & 0.98086 \\
		&&& (15.15) & (0.00434) & (0.00492) & (15.22) & (0.00435) & (0.00492) & (23.35) & (0.00462) & (0.02551) \\
		&&GP & 1.24 & 0.00000 & 0.00191 & 0.00 & 0.00000 & 0.00191 & 64.03 & 0.00000 & 0.09831 \\
		&&& (0.59) & (0.00000) & (0.00090) & (0.00) & (0.00000) & (0.00090) & (23.57) & (0.00000) & (0.03550) \\
		&&P & 676.98 & 0.04008 & 1.00000 & 676.98 & 0.04008 & 1.00000 & 672.19 & 0.03327 & 1.00000 \\
		&&& (14.52) & (0.01562) & (0.00000) & (14.52) & (0.01562) & (0.00000) & (15.21) & (0.01160) & (0.00000) \\
		&&& &&& &&& &&\\
		&$C_{2}$ & ZIGP & 643.89 & 0.01296 & 0.97781 & 634.14 & 0.00884 & 0.96730 & 612.78 & 0.00605 & 0.93724 \\
		&&& (80.65) & (0.00675) & (0.11962) & (80.75) & (0.00465) & (0.11952) & (56.92) & (0.00369) & (0.08339) \\
		&&ZIP & 657.03 & 0.01151 & 0.99950 & 656.97 & 0.01150 & 0.99950 & 654.92 & 0.01251 & 0.99522 \\
		&&& (15.49) & (0.00577) & (0.00512) & (15.66) & (0.00578) & (0.00512) & (19.28) & (0.00647) & (0.01547) \\
		&&GP & 1.24 & 0.00000 & 0.00191 & 0.00 & 0.00000 & 0.00191 & 67.67 & 0.00000 & 0.10389 \\
		&&& (0.59) & (0.00000) & (0.00090) & (0.00) & (0.00000) & (0.00090) & (25.38) & (0.00000) & (0.03816) \\
		&&P & 681.10 & 0.04588 & 1.00000 & 681.10 & 0.04588 & 1.00000 & 673.06 & 0.03455 & 1.00000 \\
		&&& (13.56) & (0.01729) & (0.00000) & (13.56) & (0.01729) & (0.00000) & (14.52) & (0.01102) & (0.00000) \\
		&&& &&& &&& &&\\
		\hline
		&&& &&& &&& &&\\
		ZIGP$_{6}$ &$C_1$&ZIGP& 307.49 & 0.01984 & 0.45465 & 307.47 & 0.01990 & 0.45465 & 383.81 & 0.01954 & 0.57293 \\
		&&& (260.59) & (0.02601) & (0.37898) & (260.61) & (0.02603) & (0.37898) & (189.41) & (0.02387) & (0.27060) \\
		&&ZIP & 702.05 & 0.10970 & 0.96127 & 592.67 & 0.04123 & 0.87349 & 594.67 & 0.04115 & 0.87694 \\
		&&& (26.90) & (0.02259) & (0.01592) & (38.38) & (0.01975) & (0.04190) & (33.72) & (0.01704) & (0.03544) \\
		&&GP & 1.37 & 0.00002 & 0.00209 & 0.19 & 0.00488 & 0.00209 & 92.38 & 0.00040 & 0.14189 \\
		&&& (2.81) & (0.00074) & (0.00416) & (2.77) & (0.01091) & (0.00416) & (27.92) & (0.00212) & (0.04198) \\
		&&P & 823.15 & 0.21052 & 1.00000 & 823.15 & 0.21052 & 1.00000 & 805.39 & 0.19304 & 1.00000 \\
		&&& (12.27) & (0.01884) & (0.00000) & (12.27) & (0.01884) & (0.00000) & (15.15) & (0.01961) & (0.00000) \\
		&&& &&& &&& &&\\
		&$C_{2}$ &ZIGP& 319.34 & 0.01029 & 0.48100 & 319.31 & 0.01029 & 0.48100 & 390.02 & 0.01116 & 0.58950 \\
		&&& (217.41) & (0.01668) & (0.32111) & (217.44) & (0.01668) & (0.32111) & (159.49) & (0.01647) & (0.23071) \\
		&&ZIP & 712.53 & 0.11166 & 0.97368 & 665.87 & 0.07683 & 0.94700 & 653.12 & 0.06752 & 0.93548 \\
		&&& (25.11) & (0.01854) & (0.01448) & (56.82) & (0.03946) & (0.04093) & (44.11) & (0.03039) & (0.03306) \\
		&&GP & 1.31 & 0.00000 & 0.00201 & 0.11 & 0.00000 & 0.00201 & 95.50 & 0.00008 & 0.14667 \\
		&&& (2.00) & (0.00000) & (0.00300) & (1.94) & (0.00000) & (0.00300) & (27.36) & (0.00095) & (0.04083) \\
		&&P & 823.43 & 0.21082 & 1.00000 & 823.43 & 0.21082 & 1.00000 & 803.25 & 0.19096 & 1.00000 \\
		&&& (12.85) & (0.01674) & (0.00000) & (12.85) & (0.01674) & (0.00000) & (15.06) & (0.01679) & (0.00000) \\
		&&& &&& &&& &&\\
		\hline
		&&& &&& &&& &&\\
		ZIP$_{3}$ &$C_1$& 	ZIGP & 650.68 & 0.00139 & 1.00000 & 636.68 & 0.00092 & 0.97882 & 619.64 & 0.00038 & 0.95323 \\
		&&& (14.59) & (0.00150) & (0.00000) & (23.85) & (0.00137) & (0.02500) & (17.52) & (0.00086) & (0.01465) \\
		&&ZIP& 650.36 & 0.00139 & 0.99952 & 650.35 & 0.00139 & 0.99952 & 636.75 & 0.00094 & 0.97899 \\
		&&& (14.88) & (0.00150) & (0.00448) & (14.91) & (0.00150) & (0.00448) & (22.60) & (0.00137) & (0.02522) \\
		&&GP & 107.04 & 0.00026 & 0.16526 & 105.11 & 0.00138 & 0.16393 & 184.21 & 0.00011 & 0.28388 \\
		&&& (239.05) & (0.00091) & (0.36896) & (237.61) & (0.00171) & (0.36605) & (195.71) & (0.00055) & (0.30250) \\
		&&P & 655.00 & 0.00783 & 1.00000 & 655.00 & 0.00783 & 1.00000 & 654.29 & 0.00689 & 1.00000 \\
		&&& (15.64) & (0.01538) & (0.00000) & (15.64) & (0.01538) & (0.00000) & (14.22) & (0.00593) & (0.00000) \\
		&&& &&& &&& &&\\
		&$C_{2}$ & 	ZIGP & 641.71 & 0.00208 & 0.98573 & 610.09 & 0.00037 & 0.93888 & 610.13 & 0.00026 & 0.93889 \\
		&&& (72.01) & (0.00256) & (0.10856) & (70.63) & (0.00094) & (0.10539) & (48.38) & (0.00069) & (0.07204) \\
		&&ZIP & 644.72 & 0.00120 & 0.99096 & 643.76 & 0.00118 & 0.99085 & 634.46 & 0.00099 & 0.97541 \\
		&&& (20.33) & (0.00147) & (0.01905) & (21.87) & (0.00147) & (0.01932) & (23.03) & (0.00164) & (0.02611) \\
		&&GP & 106.40 & 0.00026 & 0.16431 & 104.47 & 0.00136 & 0.16299 & 186.78 & 0.00011 & 0.28780 \\
		&&& (238.38) & (0.00090) & (0.36802) & (236.93) & (0.00170) & (0.36511) & (193.93) & (0.00055) & (0.29983) \\
		&&P & 660.72 & 0.01636 & 1.00000 & 660.72 & 0.01636 & 1.00000 & 657.67 & 0.01204 & 1.00000 \\
		&&& (14.82) & (0.02008) & (0.00000) & (14.82) & (0.02008) & (0.00000) & (12.48) & (0.00911) & (0.00000) \\
		&&& &&& &&& &&\\
		\hline
		\hline
	\end{tabular}
\end{center}
\vspace{5mm}
\normalsize
\newpage
\noindent
Another scenario considered is when the true non-null distribution is Geometric, the proportion of null cases is 0.85 but the fraction of zeros is 0.40.  Unlike the scenario presented in Table 1 and Figure 1, this means that the specified proportion of zeros is reduced to half.  The interest is to determine whether there would be a change in pattern should there be a significant decrease in the number of positions without a mutation.  The histograms are displayed in Figure 4 and the corresponding numerical comparisons are presented in Tables 4 and 5 found in the Supplementary section.  It can be noted that regardless of the magnitude of the fraction of zeros, a similar pattern can be observed in terms of the superiority of $C_1$ as a method for choosing $C$.  However, for the heavily mixed case presented in ZIGP$_8$, the $\widehat{FDR}$ for the two-stage procedure is slightly higher than the specified level which is 0.05.
\\\\
In addition, another scenario considered is when the true non-null distribution is Binomial($n = 250, p=0.20$), $\pi_0 = 0.70$ and $\eta$ is 0.40.  The goal is to determine whether there would be a change in pattern of results if there is a drop in the proportion of the null cases in the mixture model as compared to the results in Table 2 and Figure 2.  The histograms are displayed in Figure 5 and the numerical comparisons are presented in Tables 6 and 7 also found in the Supplementary section.  Results revealed that even with the decrease in the value of $\pi_0$, a similar pattern can be observed in terms of the superiority of $C_1$ as a method for choosing $C$ particularly for the overdispersed and heavily mixed case presented in ZIGP$_{10}$. Moreover, for ZIGP$_{10}$, Storey's procedure yielded more rejections and a higher $\widehat{TPR}$ compared to the local FDR procedure where one-stage and two-stage procedure results coincided.
\\\\
Overall, for the well-separated and moderately mixed case, if the null model is correctly specified, using the Two-Stage procedure yields $\widehat{FDR}$ closest to the nominal level $\alpha$.  Consequently, the Two-Stage procedure is superior in terms of $\widehat{TPR}$ in most cases.  If the true null model is ZIGP and the null model is correctly specified, $\widehat{FDR}$ is controlled in all procedures.  However, the Two-Stage procedure is better than the One-Stage procedure and Storey's procedure in terms of $\widehat{TPR}$.
\\\\
It can also be noted that if the true model is ZIP and ZIGP is used to model the null distribution, then the Two-Stage Procedure still yields the closest $\widehat{FDR}$ to $\alpha$ and leads to higher $\widehat{TPR}$ as compared to the other procedures.  This implies using the Two-Stage Procedure when the null model is misspecified would still produce satisfactory results.  Moreover, regardless of the shape of the non-null distribution, the Two-Stage Procedure yields better results then the other procedures.

\newpage
\normalsize
\subsection{Application to Protein Domain Data}
\vspace{5mm}
One interesting issue is identifying the position of somatic mutations, so called hotspot, on protein domains. The key question is among fixed number of positions in a single domain, which ones are significantly different from the majority. It is a novel solution for the identification of driver mutations which lead tumor progression in somatic tumor samples and recapitulates much of what is known about how protein domain families contribute to the initiation or progression of cancer.
\\\\
As an example, we analyze the mutation data which were obtained from from the tumors of 5,848 patients from The Cancer Genome Atlas (TCGA) data portal (http://tcga-data.nci.nih.gov/tcga/, Collins and Barker, 2007). These were mapped to specific positions within protein domain models to identify clusters. TCGA MAF files were obtained on July 7th, 2014 for 20 cancer types: Adrenocortical Carcinoma (ACC), Bladder Urothelial Carcinoma (BLCA), Brain Lower Grade Glioma (LGG), Breast Invasive Carcinoma (BRCA), Colon Adenocarcinoma (COAD), Glioblastoma Multiforme (GBM), Head and Neck Squamous Cell Carcinoma (HNSC), Kidney Chromophobe (KICH), Kidney Renal Clear Cell Carcinoma (KIRC), Liver Hepatocellular Carcinoma (LHIC), Lung Adenocarcinoma (LUAD), Lung Squamous Cell Carcinoma (LUSC), Ovarian Serous Cystadenocarcinoma (OV), Pancreatic Adenocarcinoma (PAAD), Prostate Adenocarcinoma (PRAD), Rectum Adenocarcinoma (READ), Skin Cutaneous Melanoma (SKCM), Stomach Adenocarcinoma (STAD), Thyroid Carcinoma (THCA), and Uterine Corpus Endometrial Carcinoma (UCEC). The mutations were mapped to proteins and domain models (Peterson et al., 2010 and Peterson et al., 2012).
\\\\
Among several hundreds of domains, we focus on five functionally well-known domains to identify the hotspots in TCGA/GBF dataset. We start with the hotspots on growth factors (cd00031), which are known to harbor reoccurring somatic mutations involved with clonal expansion, invasion across tissue barriers, and colonization of distant niches (\cite{Jeanes2008cadherins}; \cite{Takeichi1993cadherins}; \cite{Witsch2010roles}). Furthermore, protein kinases (cd00180) and the RAS-Like GTPase family of genes (cd00882), which are well-known for their role in regulating pathways important to cancer (\cite{Ahn2011map2k4}; \cite{Davies2005somatic}; \cite{Balschun2011kras}; \cite{Ohmori1997activated}; \cite{Tsatsanis2000role}). Genes with kinases or RAS-Like GTPases are expected to harbor driver mutations that reoccur at specic sites since they are classic examples of proto-oncogenes that mutate into oncogenes, contributing to cancer (\cite{Anderson1992role}; \cite{Cline1987keynote}). Additionally, we identify hotspots on ankyrin domains (cd00204), which play a role in mediating protein-protein interactions important in cancer (\cite{Li2006ankyrin}; \cite{Imaoka2014overexpression}). Furthermore, we find hotspots on transmsmbrane domains of proteins that are known to be involved with signal transduction, which is relevant in controlling processes involved with cancer (\cite{Rowinsky2003signal}; \cite{Sever2015signal}) and experimental evidence confirms the important regulatory role played by membrane proteins in cancer  (\cite{Kampen2011membrane}; \cite{Leth2009metastasis}; \cite{Neuhaus2009activation}; \cite{Sanz2014promotion}; \cite{Morita2016olfactory}; 
\newpage
\noindent
Since the mutation counts are discrete, we apply our proposed method based on various discrete models, such as Zero-Inflated Generalized Poisson, Zero-Inflated Poisson, Generalized Poisson and ordinary Poisson distribution for $f_0$. The estimated parameters based on those models are reported in Table 8 and the identified number of positions which are mutated differently from expected are in Table 9. Figure 6 shows the distribution of each protein domain and its total number of positions.
\\\\
For example, when we conduct hypothesis testing framework of section 3 to identify hotspots under the assumption of $f_0$ follows ZIGP, the results show that the identified hotspots on growth factor domain (cd00031) based on one stage and two procedures are 143 positions based on
$C_2$ among total of 366 positions. On the other hand, the local FDR with $C_1$ identifies more hotspots for two stage (201) than one stage (191) and Storey's procedure (200). Rest of domains can be analyzed in the similar manner.
\noindent
\begin{center}
\scriptsize
\textbf{Table 8.}  Comparison of Parameter Estimates for Protein Domain Data\\
\vspace{3mm}
\begin{tabular}{l|l|rrrrrr|rrrrrr}
	\hline
	\hline
	& &\multicolumn{6}{c|}{$C_1$} & \multicolumn{6}{c}{$C_2$}\\
	\hline
	Data & Model for $f_0$ & $\eta$ & $\lambda$ & $\theta$ & $\pi$ & $C$ & $D$ & $\eta$ & $\lambda$ & $\theta$ & $\pi$ & $C$ & $D$ \\
	\hline
	\texttt{cd00031}&ZIGP & 0.3246 & 1.9168 & 0.1416 & 0.4576 & 6 & 7& 0.2253 & 2.1449 & 0.5738 & 0.6139 & 4 & 36 \\
	& ZIP& 0.2289 & 1.0949 & NA & 0.3985 & 3 & 6& 0.2760 & 1.3856 & NA & 0.4244 & 2 & 6 \\
	& GP& NA & 1.5559 & 0.6609 & 0.5540 & 11 & 36& NA & 2.0944 & 0.6668 & 0.8321 & 3 & 36 \\
	& P& NA & 0.8082 & NA & 0.3944 & 3 & 6& NA & 0.7994 & NA & 0.3929 & 2 & 6 \\
	\hline
	\texttt{cd00180} &ZIGP & 0.5773 & 1.7754 & 0.2255 & 0.7095 & 7 & 10 & 0.4569 & 2.0310 & 0.7021 & 0.8716 & 5 & 63 \\
	& ZIP & 0.5507 & 1.1095 & NA & 0.6588 & 3 & 7& 0.5331 & 0.9701 & NA & 0.6484 & 2 & 7 \\
	& GP & NA & 1.3097 & 0.8292 & 0.8379 & 17 & 63& NA & 1.5161 & 0.8287 & 0.9925 & 7 & 63 \\
	& P & NA & 0.2419 & NA & 0.5864 & 1 & 7 & NA & 0.2419 & NA & 0.5864 & 1 & 7\\
	\hline
	\texttt{cd00204} &ZIGP & 0.5060 & 1.2628 & 0.0002 & 0.6853 & 5 & 6 & 0.5062 & 1.2645 & 0.0002 & 0.6854 & 4 & 5\\
	& ZIP & 0.1287 & 0.5081 & NA & 0.6784 & 3 & 7 & 0.1403 & 0.5188 & NA & 0.6792 & 2 & 7\\
	& GP & NA & 1.1923 & 0.7275 & 0.7801 & 12 & 34& NA & 1.2048 & 0.7372 & 0.7908 & 10 & 34 \\
	& P & NA & 0.4409 & NA & 0.6780 & 3 & 7& NA & 0.4089 & NA & 0.6661 & 1 & 7 \\
	\hline
	\texttt{cd00882} &ZIGP & 0.6736 & 1.3969 & 0.0003 & 0.8003 & 4 & 5 & 0.6736 & 1.3969 & 0.0003 & 0.8003 & 4 & 5\\
	& ZIP & 0.5201 & 0.6786 & NA & 0.7907 & 3 & 7 & 0.5136 & 0.6616 & NA & 0.7896 & 2 & 7\\
	& GP & NA & 1.2716 & 0.7423 & 1.0000 & 9 & 25 & NA & 1.2888 & 0.7425 & 1.0000 & 7 & 25\\
	& P & NA & 0.2174 & NA & 0.7503 & 1 & 7 & NA & 0.2174 & NA & 0.7503 & 1 & 7\\
	\hline
	\texttt{pfam00001} &ZIGP & 0.0526 & 2.4020 & 0.3839 & 0.4031 & 13 & 18 & 0.0009 & 7.5244 & 0.7562 & 1.0000 & 1 & 233\\
	& ZIP & 0.0000 & 44.9641 & NA & 1.0000 & 18 & 45& NA & 44.9641 & NA & 1.0000 & 18 & 45 \\
	& GP & NA & 2.2464 & 0.4164 & 0.4048 & 13 & 21 & NA & 4.8034 & 0.7937 & 1.0000 & 2 & 233\\
	& P & NA & 3.7966 & NA & 0.4116 & 19 & 20 & NA & 3.7966 & NA & 0.4116 & 19 & 20\\
	\hline
	\hline
\end{tabular}\\
\textbf{Table 9.}  Comparison of Number of Rejections for Protein Domain Data\\
\vspace{3mm}
\begin{tabular}{l|l|rrrr|rrrr|rrrr}
	\hline
	\hline
	\multicolumn{2}{c|}{} & \multicolumn{4}{c|}{One-Stage Procedure} & \multicolumn{4}{c|}{Two-Stage Procedure} &
	\multicolumn{4}{c}{Storey's FDR} \\
	\hline
	Data & Method & ZIGP & ZIP & GP & P & ZIGP & ZIP & GP & P  &ZIGP & ZIP & GP & P\\
	\hline
	\hline
	\texttt{cd00031}& $C_1$ & 191 & 212 & 140 & 212 & 201 & 212 & 141 & 212 & 200 & 211 & 154 & 211 \\
	&$C_2$& 143 & 205 &  16 & 212 & 143 & 205 &  17 & 212 & 162 & 204 &  85 & 211 \\
	\hline
	\texttt{cd00180} & $C_1$ & 248 & 288 &   0 & 326 & 251 & 288 &   1 & 326 & 247 & 270 &   5 & 300 \\
	& $C_2$& 63 & 288 &   0 & 326 &  63 & 288 &   1 & 326 & 122 & 284 &   0 & 300 \\
	\hline
	\texttt{cd00204} & $C_1$ & 129 & 130 &  19 & 130 & 129 & 130 &  20 & 130 & 125 & 128 &  60 & 128 \\
	& $C_2$&129 & 130 &  12 & 130 & 130 & 130 &  13 & 130 & 125 & 128 &  52 & 128 \\
	\hline
	\texttt{cd00882} & $C_1$ & 148 & 155 &   0 & 169 & 155 & 155 &   1 & 169 & 147 & 154 &   2 & 160 \\
	& $C_2$&148 & 155 &   0 & 169 & 155 & 155 &   1 & 169 & 147 & 154 &   2 & 160 \\
	\hline
	\texttt{pfam00001} & $C_1$ & 255 & 340 & 253 & 265 & 255 & 405 & 253 & 265 & 242 & 206 & 240 & 254 \\
	& $C_2$& 87 & 340 &  57 & 265 &  87 & 405 &  57 & 265 & 148 & 206 & 174 & 254 \\
	\hline
\end{tabular}
\end{center}
\normalsize
\vspace{10mm}
Results from Table 9 revealed that using $C_1$ yields more rejections.  This suggests that the data analysis for the real data shows the same pattern as the simulation results presented previously. Moreover, two domains can be highlighted in terms of the difference in the number of rejections, namely, cd00180 and pfam00001.  The number of rejections using $C_1$ is almost four times higher if the model used for $f_0$ is ZIGP and the procedure employed is either local FDR or two-stage method.  Using Storey's FDR, the number of rejections using $C_1$ is almost twice given that the model for $f_0$ is ZIGP.  Overall, the results for the real data analysis is consistent with the simulation studies.
\newpage
\noindent
\scriptsize
\begin{center}
\begin{tabular}{ccc}
	\begin{tabular}{c}
		\includegraphics[width=0.3\linewidth]{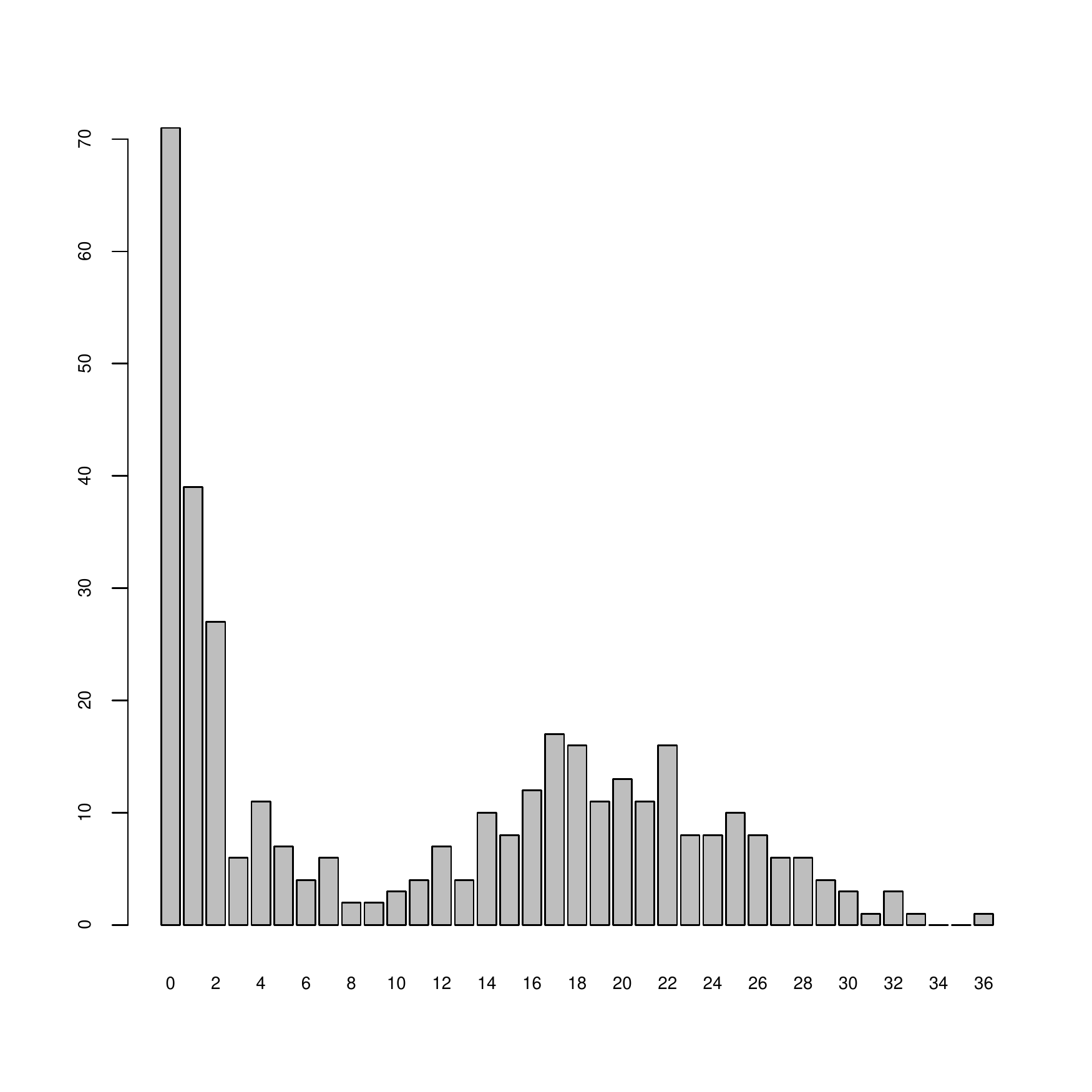}\\
		cd00031 $N = 366$\\
	\end{tabular}
	& \begin{tabular}{c}
		\includegraphics[width=0.3\linewidth]{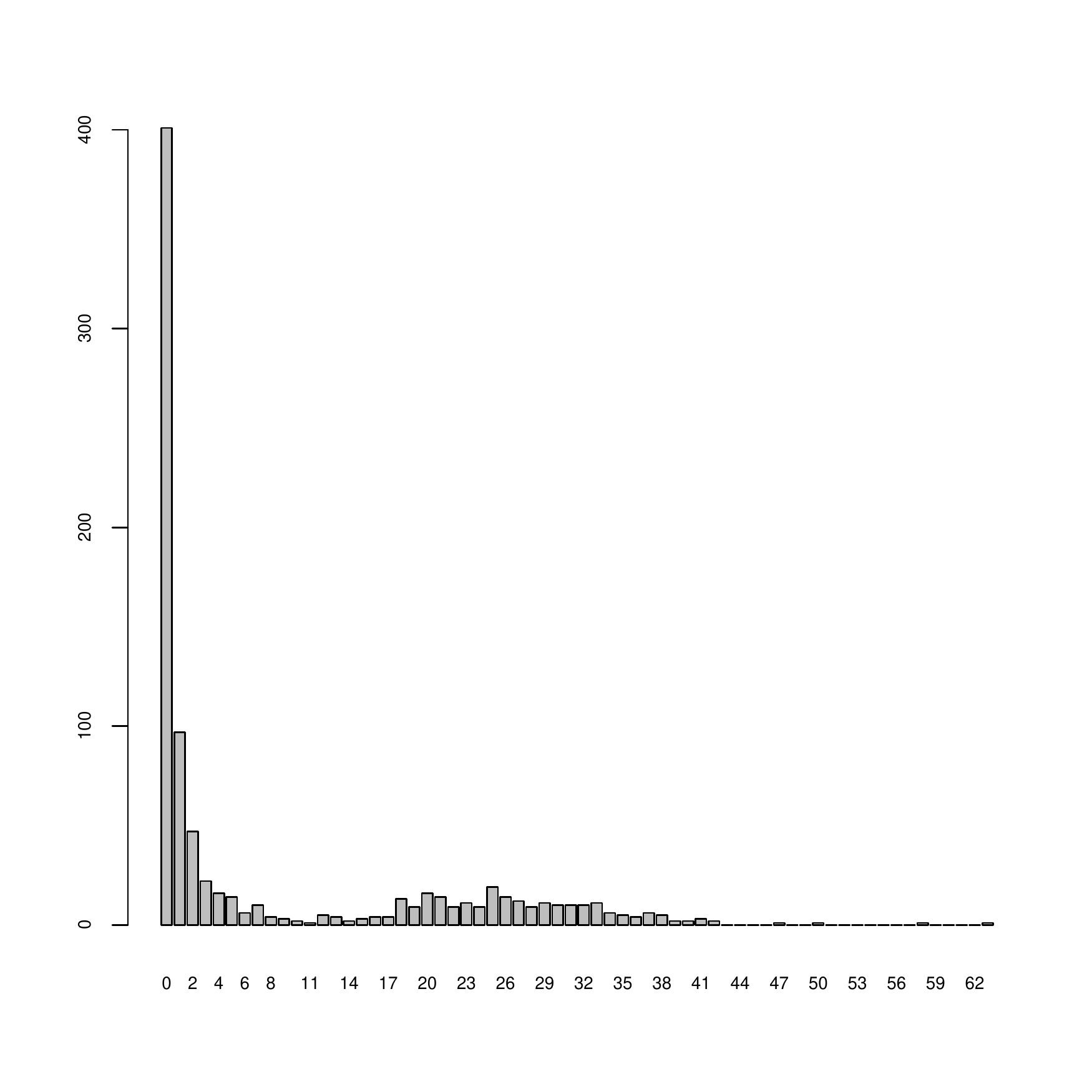}\\
		cd00180 $N = 871$\\
	\end{tabular}
	& \begin{tabular}{c}
		\includegraphics[width=0.3\linewidth]{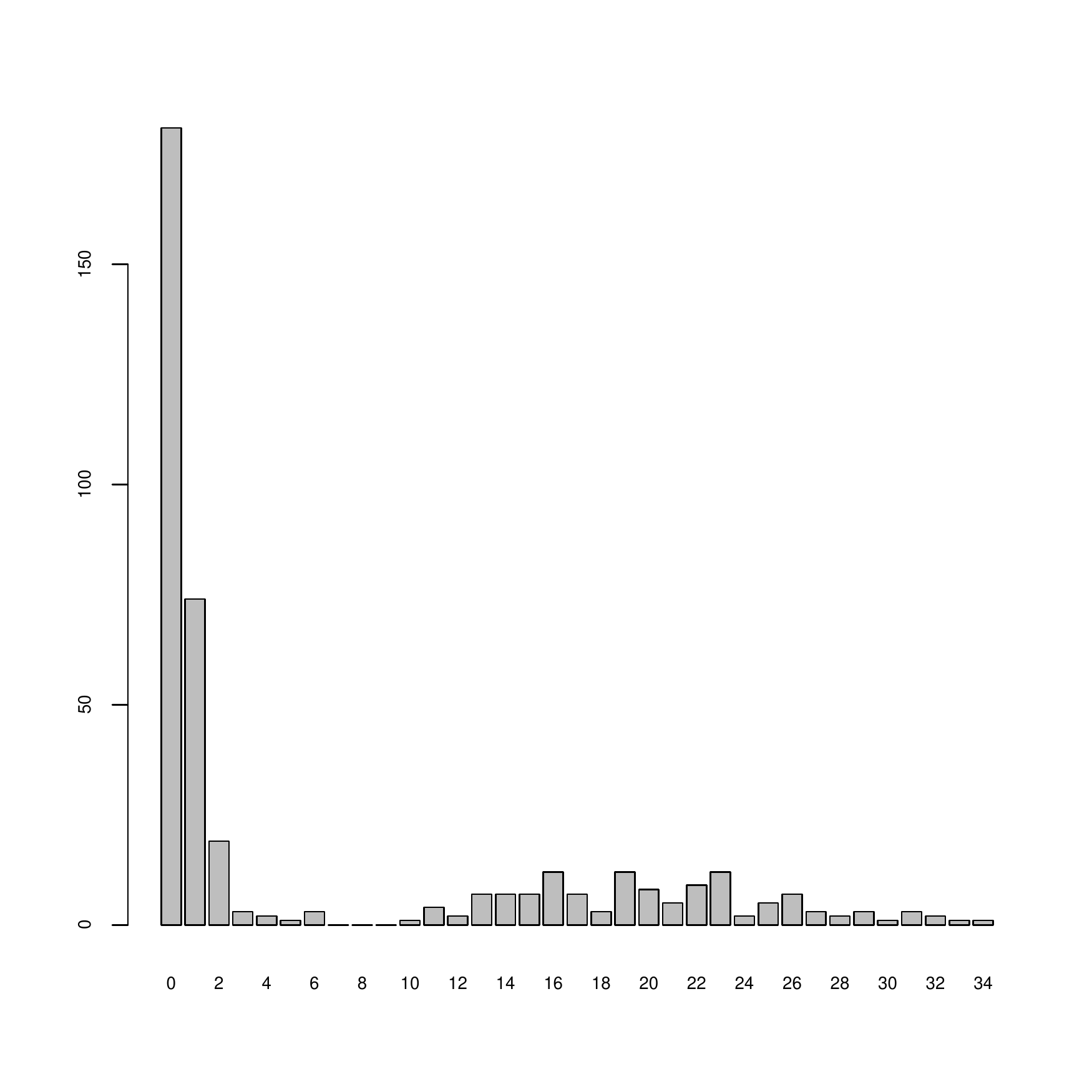}\\
		cd00204 $N = 409$\\
	\end{tabular}\\
\end{tabular}
\begin{tabular}{cc}
	\begin{tabular}{c}
		\includegraphics[width=0.3\linewidth]{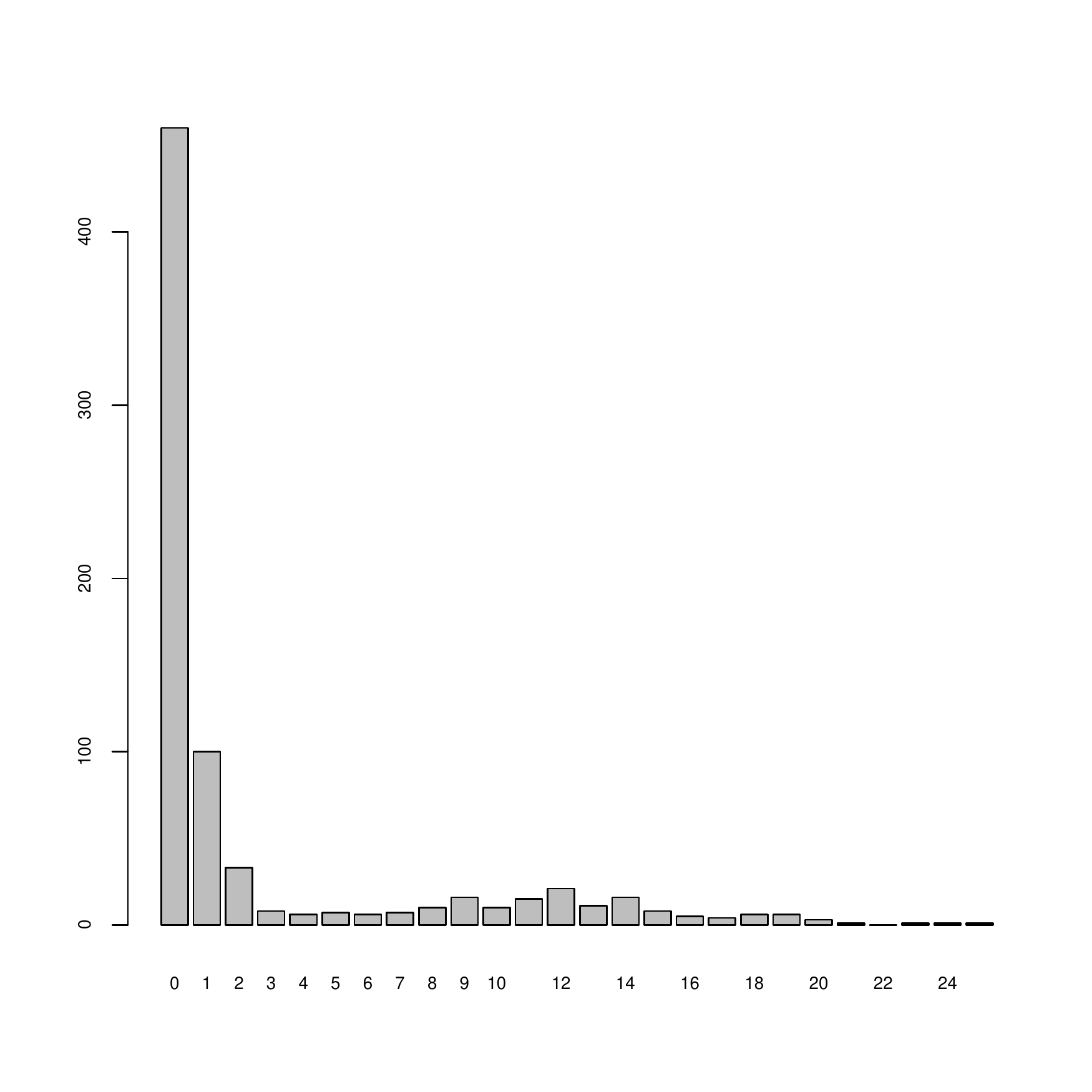}\\
		cd00882 $N = 762$\\
	\end{tabular}
	& \begin{tabular}{c}
		\includegraphics[width=0.3\linewidth]{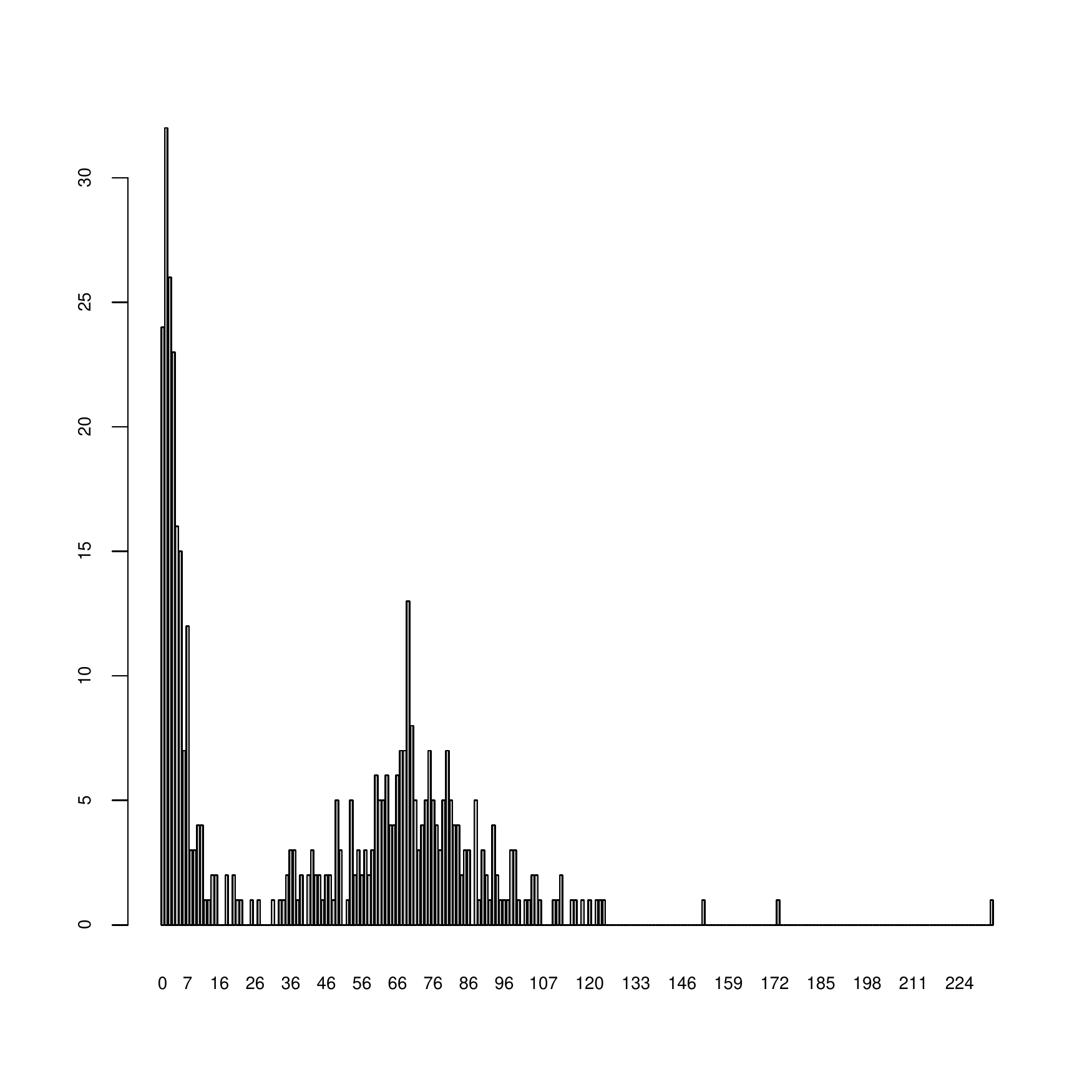}\\
		pfam00001 $N = 430$\\
	\end{tabular}
\end{tabular}\\
\vspace{5mm}
\textbf{Figure 6.}  Histogram of Protein Domain Data\\
\end{center}

\vspace{10mm}
\normalsize
\section{Conclusion}
\noindent In this paper, our main interest is to select significant mutation counts while controlling a given level of Type I error via False Discovery Rate (FDR) procedures. We assume that if the number of mutations $a \leq C$, then $a$ is guaranteed to be from the null model, for some positive integer $C$.  We propose a method for identify a cut-off $C$ and show that this is superior to the cut-off developed by extending Efron's proposal.  In addition, after the selection of this cut-off, we consider a screening process so that the number of mutations exceeding a certain value $D$ $(D > C)$ should be considered as significant mutations.
This two-stage procedure in the selection of $C$ and $D$ yielded a testing procedure with increased power compared to Efron's local FDR and Storey's FDR particularly if the non-null distribution behaves similarly to the Geometric distribution and if the null distribution is well-separated and overdispersion is not observed.

\newpage
\normalsize
\bibliography{literature}
\bibliographystyle{plain}
\newpage

\section*{Appendix:}
\noindent \large \textbf{E- Step:}\\
\\
\normalsize
At the $(p + 1)th$ stage, the expectation $Q(\Theta; \Theta^{(p)})$ of the log-likelihood of the complete data specified in \eqref{eqn:llh} can be computed conditional on the observed data $\boldsymbol{y}_n$ and the current fit $\Theta^{(p)}$ for $\Theta$.
\begin{eqnarray*}\label{eqn:expllh}
	Q(\Theta; \Theta^{(p)}) &=& n_0 \tau_{00}(\Theta^{(p)})\log \eta +  \sum \limits_{j = 0}^{C} n_j \tau_{1j}(\Theta^{(p)}) \log (1 - \eta) + \sum \limits_{j = 0}^{C} n p_j (\Theta^{(p)}) \tau_{1j}(\Theta^{(p)}) \log (1 - \eta)\\
	&&+ (\log \lambda - \lambda) \sum \limits_{j = 0}^{C} n_j \tau_{1j}(\Theta^{(p)}) + (\log \lambda - \lambda) \sum \limits_{j = C+1}^{K} n p_j (\Theta^{(p)}) \tau_{1j}(\Theta^{(p)}) \\
	&&+ \sum \limits_{j = 0}^{C} n_j (j - 1) \tau_{1j}(\Theta^{(p)})\log (\lambda + \theta j) + \sum \limits_{j = C+1}^{K} n (j - 1) p_j(\Theta^{(p)})\tau_{1j}(\Theta^{(p)})\log (\lambda + \theta j)\\
	&& - \left[\theta \sum \limits_{j = 0}^{C} j n_j \tau_{1j}(\Theta^{(p)}) + \theta \sum \limits_{j = C+1}^{K} j np_j(\Theta^{(p)}) \tau_{1j}(\Theta^{(p)}) + constant\right]
\end{eqnarray*}

\vspace{5mm}
\noindent \large \textbf{M- Step:}\\
\\
\normalsize
In order to arrive at an estimate of $\Theta^{(p+1)}$ at the $(p + 1)th$ stage, the goal is to maximize $Q(\Theta; \Theta^{(p)})$ with respect to $\Theta$. The estimates of $\eta, \lambda$ and $\theta$ obtained at the $(p + 1)th$ stage are as follows:
\begin{eqnarray*}
	\eta^{(p + 1)} &=& \displaystyle \frac{n_0 \tau_{00} (\Theta^{(p)})}{n_0 \tau_{00} (\Theta^{(p)}) + \displaystyle \sum \limits_{j = 0}^{C} n_j \tau_{1j}(\Theta^{(p)}) + \sum \limits_{j = C+1}^{K} n p_j (\Theta^{(p)}) \tau_{1j}(\Theta^{(p)})}
\end{eqnarray*}
\begin{eqnarray*}
	\lambda^{(p + 1)} &=& \displaystyle \frac{\displaystyle \sum \limits_{j = 0}^{C} n_j [\tau_{1j}(\Theta^{(p)}) + (j - 1) \tau_{2j}(\Theta^{(p)})] + \sum \limits_{j = C+1}^{K} n p_j (\Theta^{(p)}) [\tau_{1j}(\Theta^{(p)}) + (j - 1) \tau_{2j}(\Theta^{(p)})]}{\displaystyle \sum \limits_{j = 0}^{C} n_j \tau_{1j}(\Theta^{(p)}) + \sum \limits_{j = C+1}^{K} n p_j (\Theta^{(p)}) \tau_{1j}(\Theta^{(p)})}\\
	\theta^{(p + 1)} &=& \displaystyle \frac{\displaystyle \sum \limits_{j = 0}^{C} n_j(j - 1) \tau_{3j}(\Theta^{(p)}) + \sum \limits_{j = C+1}^{K} n p_j(j - 1) \tau_{3j}(\Theta^{(p)})}{\displaystyle \sum \limits_{j = 0}^{C} j n_j \tau_{1j}(\Theta^{(p)}) + \sum \limits_{j = C+1}^{K} j n p_j (\Theta^{(p)}) \tau_{1j}(\Theta^{(p)})}
\end{eqnarray*}
where $\tau_{2j}(\Theta^{(p)}) = \displaystyle \frac{\lambda^{(p)}}{\lambda^{(p)} + \theta^{(p)} j}$ and $\tau_{3j}(\Theta^{(p)}) = \displaystyle \frac{\theta^{(p)} j}{\lambda^{(p)} + \theta^{(p)} j}$.
\\\\
\noindent If the null distribution is modeled using Zero-Inflated Poisson distribution then the log likelihood $\ell(\eta, \lambda \mid \boldsymbol{x}_N)$ of the entire data vector is
\begin{equation*}\label{eqn:zipllh}
n_0 \log \left(\eta + (1 - \eta)e^{-\lambda}\right) + \displaystyle \sum \limits_{j = 1}^{C} n_j \log (1 - \eta) \hspace{1.5mm} \displaystyle \frac {\lambda^j e^{-\lambda}}{j!} + \displaystyle \sum \limits_{j = C + 1}^{K} {n_j \log f(j; \cdot)}
\end{equation*}
Following the same procedure, the $E-$Step at the $(p + 1)th$ stage would yield
\begin{eqnarray*}\label{eqn:expllh}
	Q(\Theta; \Theta^{(p)}) &=& n_0 \tau_{00}(\Theta^{(p)})\log \eta +  \sum \limits_{j = 0}^{C} n_j \tau_{1j}(\Theta^{(p)}) \log (1 - \eta) + \sum \limits_{j = 0}^{C} n p_j (\Theta^{(p)}) \tau_{1j}(\Theta^{(p)}) \log (1 - \eta)\\
	&&+ \log \lambda \sum \limits_{j = 0}^{C} j n_j \tau_{1j}(\Theta^{(p)}) + \log \lambda \sum \limits_{j = C+1}^{K} j n p_j (\Theta^{(p)}) \tau_{1j}(\Theta^{(p)}) \\
	&& - \left[\lambda \sum \limits_{j = 0}^{C} n_j \tau_{1j}(\Theta^{(p)}) + \lambda \sum \limits_{j = C+1}^{K} n p_j(\Theta^{(p)}) \tau_{1j}(\Theta^{(p)}) + constant\right]
\end{eqnarray*}
For the $M-$Step, the estimates of $\eta$ and $\lambda$ obtained at the $(p + 1)th$ stage are as follows:

\begin{eqnarray*}
	\eta^{(p + 1)} &=& \displaystyle \frac{n_0 \tau_{00} (\Theta^{(p)})}{n_0 \tau_{00} (\Theta^{(p)}) + \displaystyle \sum \limits_{j = 0}^{C} n_j \tau_{1j}(\Theta^{(p)}) + \sum \limits_{j = C+1}^{K} n p_j (\Theta^{(p)}) \tau_{1j}(\Theta^{(p)})}\\
	\lambda^{(p + 1)} &=& \displaystyle \frac{\displaystyle \sum \limits_{j = 0}^{C} j n_j \tau_{1j}(\Theta^{(p)}) + \sum \limits_{j = C+1}^{K} j n p_j (\Theta^{(p)}) \tau_{1j}(\Theta^{(p)}) }{\displaystyle \sum \limits_{j = 0}^{C} n_j \tau_{1j}(\Theta^{(p)}) + \sum \limits_{j = C+1}^{K} n p_j (\Theta^{(p)}) \tau_{1j}(\Theta^{(p)})}
\end{eqnarray*}

\noindent If the null distribution is modeled using Generalized Poisson distribution then the log likelihood $\ell(\lambda, \theta \mid \boldsymbol{x}_N)$ of the entire data is
\begin{equation*}\label{eqn:gpllh}
\displaystyle \sum \limits_{j = 0}^{C} n_j \hspace{1.5mm}\log \left(\displaystyle \frac {\lambda(\lambda + \theta j)^{j - 1} e^{-\lambda - \theta j}}{j!}\right) + \displaystyle \sum \limits_{j = C + 1}^{K} {n_j \log f(j; \cdot)}
\end{equation*}
Unlike ZIGP, this model is not a mixture density so the procedure does not require the inclusion of latent variables.  The $E-$Step would yield
\begin{eqnarray*}\label{eqn:gpexpllh}
	Q(\Theta; \Theta^{(p)}) &=& (\log \lambda - \lambda) \sum \limits_{j = 0}^{C} n_j + (\log \lambda - \lambda) \sum \limits_{j = C+1}^{K} n p_j (\Theta^{(p)}) + \sum \limits_{j = 0}^{C} n_j (j - 1) \log (\lambda + \theta j)\\
	&& + \sum \limits_{j = C+1}^{K} n (j - 1) p_j(\Theta^{(p)})\log (\lambda + \theta j)
	- \left[\theta \sum \limits_{j = 0}^{C} j n_j + \theta \sum \limits_{j = C+1}^{K} j np_j(\Theta^{(p)}) + constant\right]
\end{eqnarray*}
\\\\
\noindent At the $(p + 1)th$ stage, the $M-$Step would yield the estimates of $\lambda$ and $\theta$ as follows:
\begin{eqnarray*}
	\lambda^{(p + 1)} &=& \displaystyle \frac{\displaystyle \sum \limits_{j = 0}^{C} n_j[1 + (j - 1)\tau_{2j}(\Theta^{(p)})] + \sum \limits_{j = C+1}^{K} n p_j (\Theta^{(p)})[1 + (j - 1)\tau_{2j}(\Theta^{(p)})]}{\displaystyle \sum \limits_{j = 0}^{C} n_j + \sum \limits_{j = C+1}^{K} n p_j (\Theta^{(p)})}\\
	\theta^{(p + 1)} &=& \displaystyle \frac{\displaystyle \sum \limits_{j = 0}^{C} n_j (j - 1)\tau_{3j}(\Theta^{(p)}) + \sum \limits_{j = C+1}^{K} n (j - 1) p_j (\Theta^{(p)}) \tau_{3j}(\Theta^{(p)})}{\displaystyle \sum \limits_{j = 0}^{C} j n_j + \sum \limits_{j = C+1}^{K} j n p_j (\Theta^{(p)}) }
\end{eqnarray*}
\noindent Lastly, if $f_0$ is modeled using Poisson distribution then the log likelihood $\ell(\lambda \mid \boldsymbol{x}_N)$ of the entire data vector is
\begin{equation*}\label{eqn:pllh}
\displaystyle \sum \limits_{j = 0}^{C} n_j \hspace{1.5mm}\log \left(\displaystyle \frac {\lambda^{j} e^{-\lambda}}{j!}\right) + \displaystyle \sum \limits_{j = C + 1}^{K} {n_j \log f(j; \cdot)}
\end{equation*}
Since this model is also not a mixture density then the procedure does not require the inclusion of zero-one indicator variables.  The $E-$Step would yield
\begin{eqnarray*}\label{eqn:expppllh}
	Q(\Theta; \Theta^{(p)}) &=& \log \lambda \sum \limits_{j = 0}^{C} j n_j + \log \lambda \sum \limits_{j = C+1}^{K} j n p_j (\Theta^{(p)}) \\
	&& - \left[\lambda \sum \limits_{j = 0}^{C} n_j \tau_{1j}(\Theta^{(p)}) + \lambda \sum \limits_{j = C+1}^{K} n p_j \tau_{1j}(\Theta^{(p)}) + constant\right]
\end{eqnarray*}

\noindent The $M-$Step would yield the estimate of $\lambda$  at the $(p + 1)th$ stage as follows:
\begin{eqnarray*}
	\lambda^{(p + 1)} &=& \displaystyle \frac{\displaystyle \sum \limits_{j = 0}^{C} j n_j + \sum \limits_{j = C+1}^{K} j n p_j (\Theta^{(p)})}{\displaystyle \sum \limits_{j = 0}^{C} n_j + \sum \limits_{j = C+1}^{K} n p_j (\Theta^{(p)})}
\end{eqnarray*}
\newpage
\section*{Supplementary Figures and Tables:}

\scriptsize
\begin{center}
	\begin{tabular}{ccc}	
		\begin{tabular}{c}
			\includegraphics[width=0.3\linewidth]{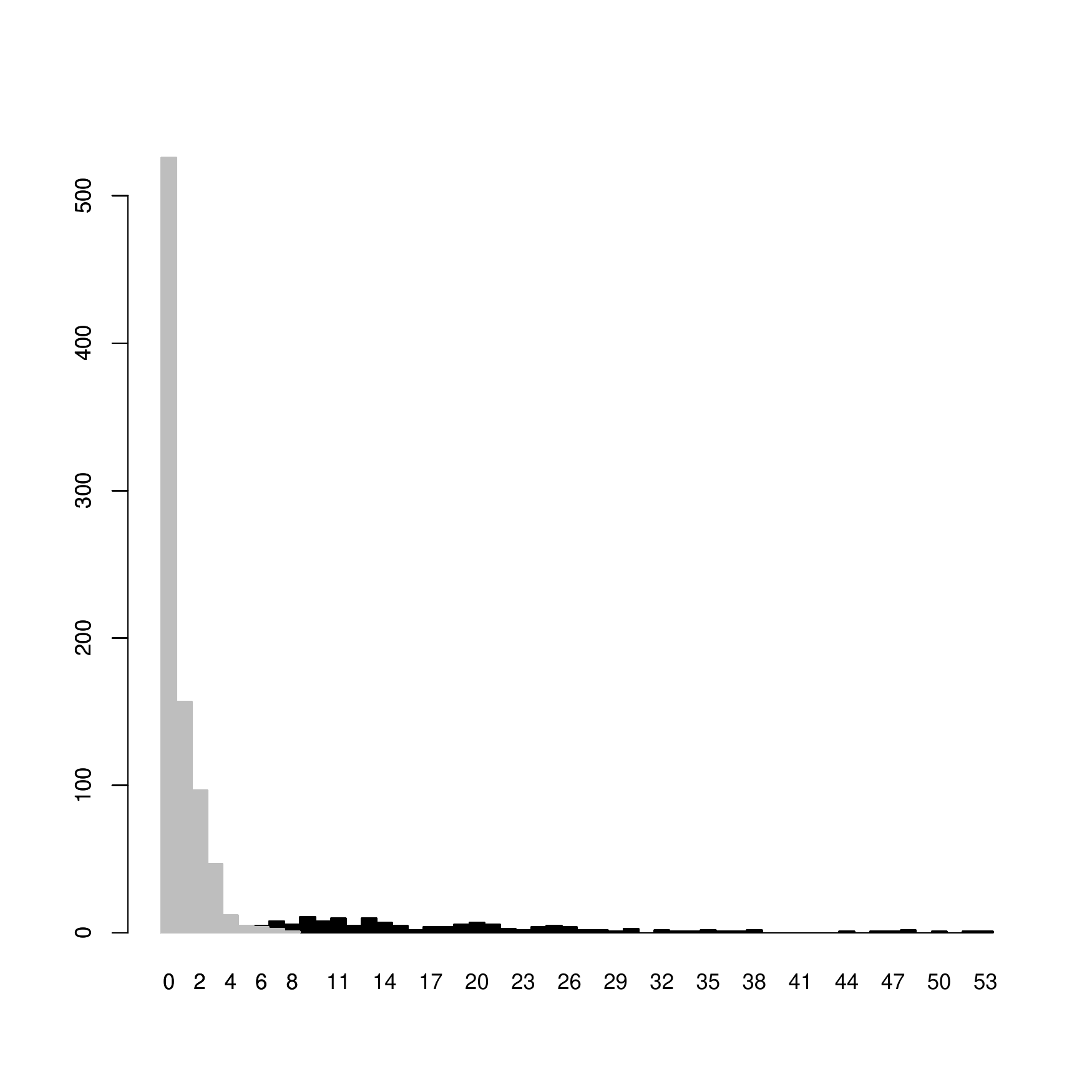}\\
			\textbf{ZIGP}$_7$\\
			ZIGP($\eta = 0.40, \lambda = 1, \theta=0.15$)\\
		\end{tabular}
		& \begin{tabular}{c}
			\includegraphics[width=0.3\linewidth]{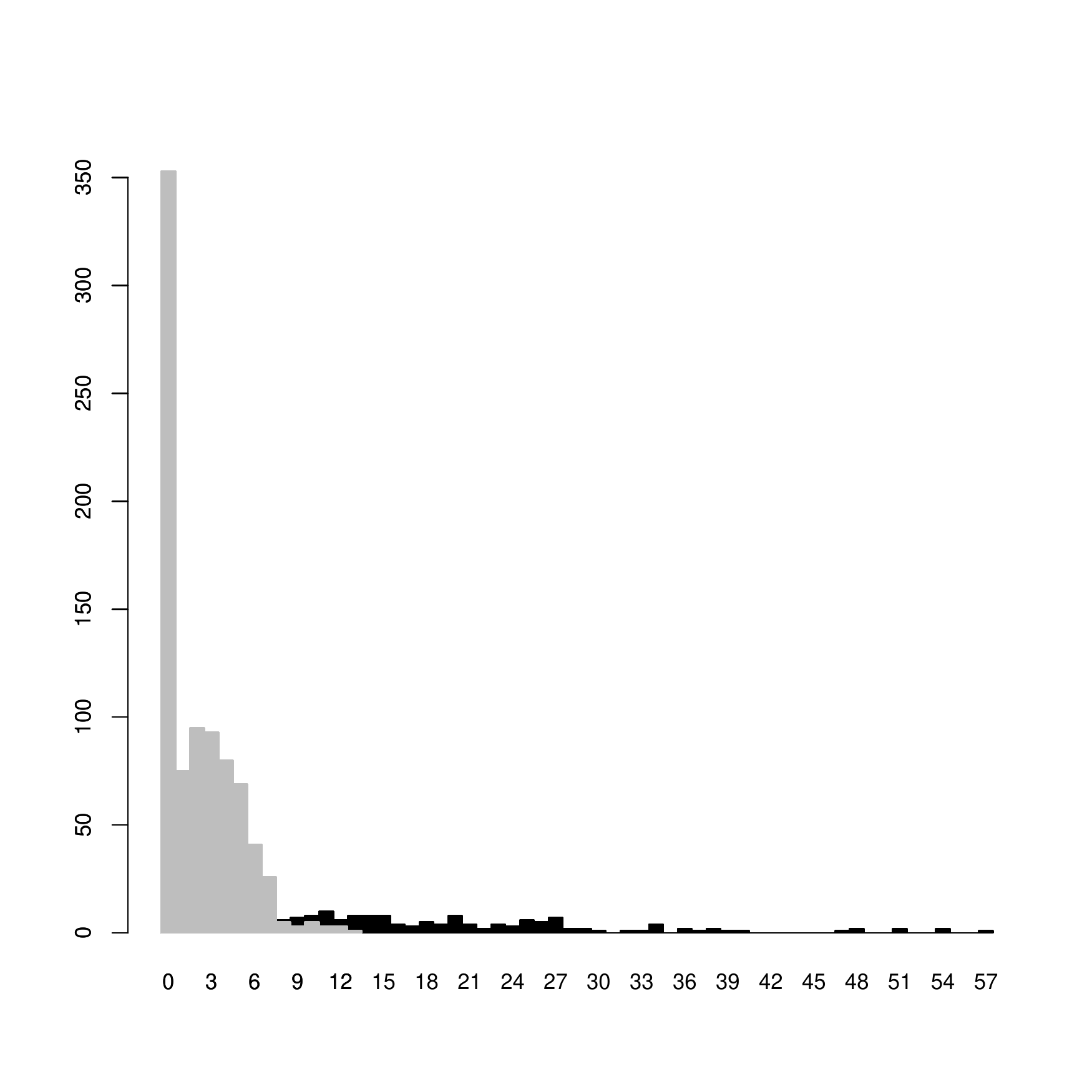}\\
			\textbf{ZIGP}$_8$\\
			ZIGP($\eta = 0.40, \lambda = 3, \theta = 0.15$)\\
		\end{tabular}
		& \begin{tabular}{c}
			\includegraphics[width=0.3\linewidth]{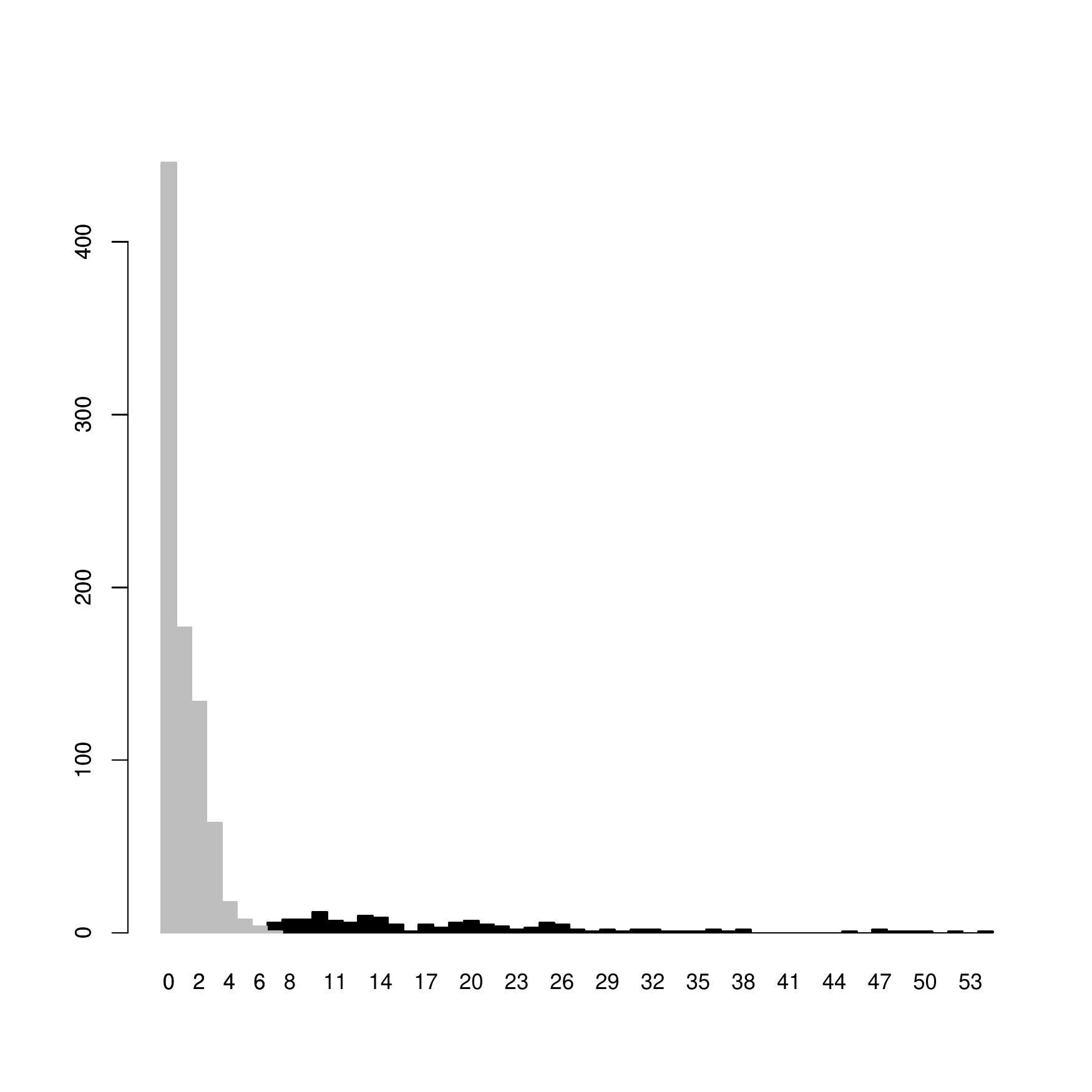}\\
			\textbf{ZIP}$_4$\\
			ZIP($\eta = 0.40, \lambda = 1.5$)\\
		\end{tabular}\\
	\end{tabular}\\
	\vspace{5mm}
	\textbf{Figure 4.} Histogram when the Non-null Distribution is Geometric($p=0.08$) and $\pi_0 = 0.85$
	
\end{center}
\vspace{10mm}
\begin{center}
	\textbf{Table 4.}  Numerical Comparison using $C_1$ as cut-off when the Non-null Distribution is Geometric($p=0.08$), $\pi_0 = 0.85$ and $\alpha = 0.05$.  The number in $(\cdot)$ represents the standard deviation.\\
	\vspace{3mm}
	\begin{tabular}{ll|ccc|ccc|ccc}
		\hline
		\hline
		& & \multicolumn{3}{c|}{Two-Stage Procedure} & \multicolumn{3}{c|}{One-Stage Procedure} &
		\multicolumn{3}{c}{Storey's FDR} \\
		\hline
		Null & Model & & & &  & & &\\
		Distribution &  for $f_0$ & $R$ & $\widehat{FDR}$ & $\widehat{TPR}$ & $R$ & $\widehat{FDR}$ & $\widehat{TPR}$ & $R$ & $\widehat{FDR}$ & $\widehat{TPR}$ \\
		\hline
		&& &&& &&& &&\\
		ZIGP$_7$ & 	ZIGP & 156.30 & 0.03821 & 0.99995 & 138.95 & 0.00700 & 0.91765 & 134.85 & 0.00440 & 0.89345 \\
		& & (12.02) & (0.02781) & (0.00091) & (12.70) & (0.00784) & (0.03306) & (11.25) & (0.00596) & (0.03003) \\
		&ZIP & 151.22 & 0.02186 & 0.98405 & 151.20 & 0.02185 & 0.98399 & 148.77 & 0.01788 & 0.97222 \\
		& & (12.36) & (0.01488) & (0.02298) & (12.37) & (0.01489) & (0.02307) & (12.26) & (0.01366) & (0.02825) \\
		&GP & 86.53 & 0.01001 & 0.56913 & 75.80 & 0.00381 & 0.50808 & 75.54 & 0.00199 & 0.50418 \\
		& & (71.46) & (0.01464) & (0.46522) & (63.43) & (0.00672) & (0.41601) & (61.58) & (0.00471) & (0.40734) \\
		&P & 190.45 & 0.20303 & 1.00000 & 190.45 & 0.20303 & 1.00000 & 175.45 & 0.14214 & 1.00000 \\
		& & (23.92) & (0.08036) & (0.00000) & (23.92) & (0.08036) & (0.00000) & (14.11) & (0.04425) & (0.00000) \\
		&& &&& &&& &&\\
		\hline
		&& &&& &&& &&\\
		ZIGP$_8$ &ZIGP& 128.65 & 0.05204 & 0.80943 & 112.85 & 0.02111 & 0.79575 & 113.47 & 0.02096 & 0.73885 \\
		& & (21.91) & (0.03398) & (0.11523) & (20.03) & (0.01892) & (0.11091) & (18.41) & (0.01722) & (0.10381) \\
		&ZIP & 195.05 & 0.25656 & 0.96497 & 142.60 & 0.08532 & 0.87498 & 135.85 & 0.06452 & 0.84347 \\
		& & (12.87) & (0.03248) & (0.01509) & (21.31) & (0.05996) & (0.05326) & (15.46) & (0.03620) & (0.04738) \\
		&GP & 2.07 & 0.00000 & 0.01419 & 1.45 & 0.00000 & 0.01419 & 10.67 & 0.00000 & 0.07234 \\
		& & (2.49) & (0.00000) & (0.01797) & (2.78) & (0.00000) & (0.01797) & (5.91) & (0.00000) & (0.04276) \\
		&P & 496.36 & 0.69548 & 1.00000 & 496.36 & 0.69548 & 1.00000 & 391.38 & 0.61397 & 1.00000 \\
		& & (42.85) & (0.03060) & (0.00000) & (42.85) & (0.03060) & (0.00000) & (35.81) & (0.03580) & (0.00000) \\
		&& &&& &&& &&\\
		\hline
		&& &&& &&& &&\\
		ZIP$_4$ & 	ZIGP & 152.37 & 0.01508 & 0.99869 & 135.25 & 0.00055 & 0.89951 & 135.46 & 0.00056 & 0.90103 \\
		& & (11.56) & (0.01184) & (0.00516) & (11.01) & (0.00203) & (0.02502) & (10.87) & (0.00204) & (0.02431) \\
		&ZIP & 144.35 & 0.00341 & 0.95734 & 142.47 & 0.00288 & 0.94463 & 141.56 & 0.00247 & 0.93979 \\
		& & (11.37) & (0.00527) & (0.01708) & (12.02) & (0.00507) & (0.02194) & (11.73) & (0.00438) & (0.03154) \\
		&GP & 146.60 & 0.00976 & 0.96605 & 128.21 & 0.00037 & 0.85250 & 126.81 & 0.00030 & 0.84407 \\
		& & (17.35) & (0.01082) & (0.08442) & (16.72) & (0.00168) & (0.08619) & (16.29) & (0.00153) & (0.08847) \\
		&P & 197.73 & 0.22736 & 1.00000 & 197.73 & 0.22736 & 1.00000 & 163.45 & 0.07836 & 1.00000 \\
		& & (30.67) & (0.09507) & (0.00000) & (30.67) & (0.09507) & (0.00000) & (14.77) & (0.04829) & (0.00000) \\
		&& &&& &&& &&\\
		\hline
		\hline
	\end{tabular}
\end{center}
\newpage

\begin{center}
	\textbf{Table 5.}  Numerical Comparison using $C_2$ as cut-off, when the Non-null Distribution is Geometric($p=0.08$), $\pi_0 = 0.85$ and $\alpha = 0.05$.  The number in $(\cdot)$ represents the standard deviation.\\
	\vspace{3mm}
	\begin{tabular}{ll|ccc|ccc|ccc}
		\hline
		\hline
		& & \multicolumn{3}{c|}{Two-Stage Procedure} & \multicolumn{3}{c|}{One-Stage Procedure} &
		\multicolumn{3}{c}{Storey's FDR} \\
		\hline
		Null & Model & & & &  & & &\\
		Distribution &  for $f_0$ & $R$ & $\widehat{FDR}$ & $\widehat{TPR}$ & $R$ & $\widehat{FDR}$ & $\widehat{TPR}$ & $R$ & $\widehat{FDR}$ & $\widehat{TPR}$ \\
		\hline
		&& &&& &&& &&\\
		ZIGP$_7$ & 	ZIGP & 158.52 & 0.05181 & 1.00000 & 134.86 & 0.00455 & 0.89319 & 134.00 & 0.00395 & 0.88832 \\
		& & (11.85) & (0.02711) & (0.00000) & (11.49) & (0.00627) & (0.02918) & (10.86) & (0.00550) & (0.02645) \\
		&ZIP & 151.00 & 0.02163 & 0.98292 & 150.94 & 0.02158 & 0.98264 & 148.78 & 0.01803 & 0.97213 \\
		& & (12.33) & (0.01490) & (0.02383) & (12.38) & (0.01494) & (0.02437) & (12.22) & (0.01396) & (0.02868) \\
		&GP & 86.48 & 0.01003 & 0.56873 & 75.74 & 0.00378 & 0.50698 & 75.15 & 0.00220 & 0.50151 \\
		& & (71.43) & (0.01467) & (0.46501) & (63.40) & (0.00670) & (0.41646) & (61.91) & (0.00489) & (0.40962) \\
		&P & 190.45 & 0.20303 & 1.00000 & 190.45 & 0.20303 & 1.00000 & 175.45 & 0.14214 & 1.00000 \\
		& & (23.92) & (0.08036) & (0.00000) & (23.92) & (0.08036) & (0.00000) & (14.11) & (0.04425) & (0.00000) \\ 		
		&& &&& &&& &&\\
		\hline
		&& &&& &&& &&\\
		ZIGP$_8$ &ZIGP& 91.67 & 0.03038 & 0.58328 & 80.33 & 0.01183 & 0.57384 & 83.39 & 0.01184 & 0.54606 \\
		& & (56.87) & (0.03400) & (0.35421) & (49.98) & (0.01687) & (0.34786) & (46.34) & (0.01584) & (0.29789) \\
		&ZIP & 195.05 & 0.25656 & 0.96497 & 136.32 & 0.07661 & 0.83890 & 130.62 & 0.05678 & 0.81515 \\
		& & (12.87) & (0.03248) & (0.01509) & (27.85) & (0.06723) & (0.10124) & (20.64) & (0.04213) & (0.08111) \\
		&GP & 1.05 & 0.00000 & 0.00703 & 0.00 & 0.00000 & 0.00703 & 4.48 & 0.00000 & 0.02989 \\
		& & (0.23) & (0.00000) & (0.00162) & (0.00) & (0.00000) & (0.00162) & (1.58) & (0.00000) & (0.01052) \\
		&P & 496.36 & 0.69548 & 1.00000 & 496.36 & 0.69548 & 1.00000 & 391.38 & 0.61397 & 1.00000 \\
		& & (42.85) & (0.03060) & (0.00000) & (42.85) & (0.03060) & (0.00000) & (35.81) & (0.03580) & (0.00000) \\
		&& &&& &&& &&\\
		\hline
		&& &&& &&& &&\\
		ZIP$_4$ & 	ZIGP & 152.73 & 0.01742 & 0.99870 & 131.27 & 0.00036 & 0.87284 & 129.89 & 0.00032 & 0.86435 \\
		& & (11.44) & (0.01475) & (0.00789) & (12.02) & (0.00159) & (0.03779) & (11.40) & (0.00152) & (0.04224) \\
		&ZIP & 144.35 & 0.00341 & 0.95734 & 140.70 & 0.00234 & 0.93476 & 139.93 & 0.00197 & 0.92932 \\
		& & (11.37) & (0.00527) & (0.01708) & (12.46) & (0.00471) & (0.03069) & (12.11) & (0.00391) & (0.03658) \\
		&GP & 147.50 & 0.01106 & 0.97085 & 127.70 & 0.00036 & 0.84953 & 126.42 & 0.00029 & 0.84149 \\
		& & (17.46) & (0.01080) & (0.08649) & (16.84) & (0.00166) & (0.08692) & (16.41) & (0.00151) & (0.08957) \\
		&P & 197.73 & 0.22736 & 1.00000 & 197.73 & 0.22736 & 1.00000 & 163.45 & 0.07836 & 1.00000 \\
		& & (30.67) & (0.09507) & (0.00000) & (30.67) & (0.09507) & (0.00000) & (14.77) & (0.04829) & (0.00000) \\
		&& &&& &&& &&\\
		\hline
		\hline
	\end{tabular}
\end{center}
\vspace{5mm}
\newpage
\noindent
\begin{center}
	\textbf{Figure 5.} Histogram when the Non-null Distribution is Binomial($n=250, p=0.20$) and $\pi_0 = 0.70$
	\begin{tabular}{ccc}
		\begin{tabular}{c}
			\includegraphics[width=0.3\linewidth]{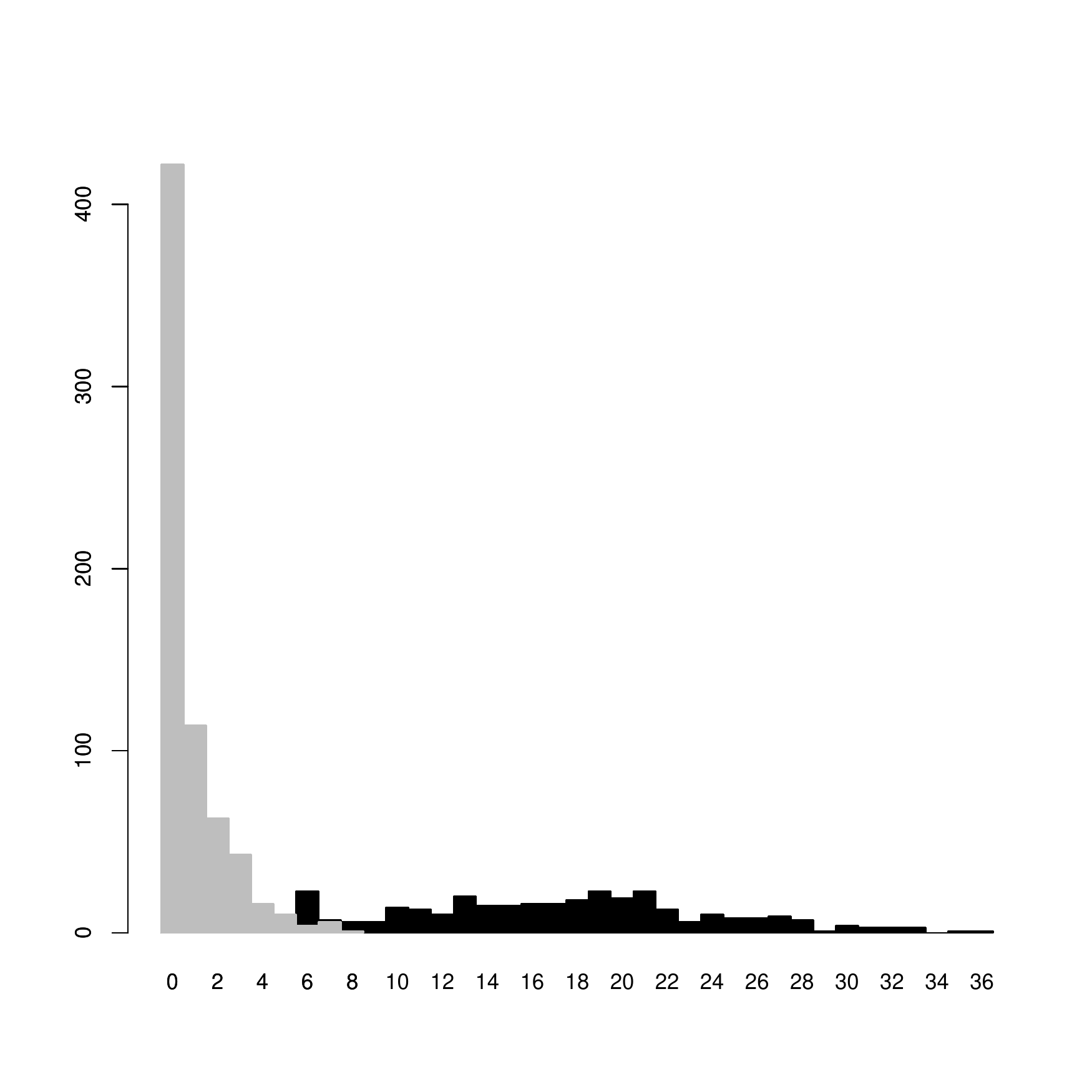}\\
			\textbf{ZIGP}$_9$\\
			ZIGP($\eta = 0.40, \lambda = 1, \theta=0.30$)\\
		\end{tabular}
		& \begin{tabular}{c}
			\includegraphics[width=0.3\linewidth]{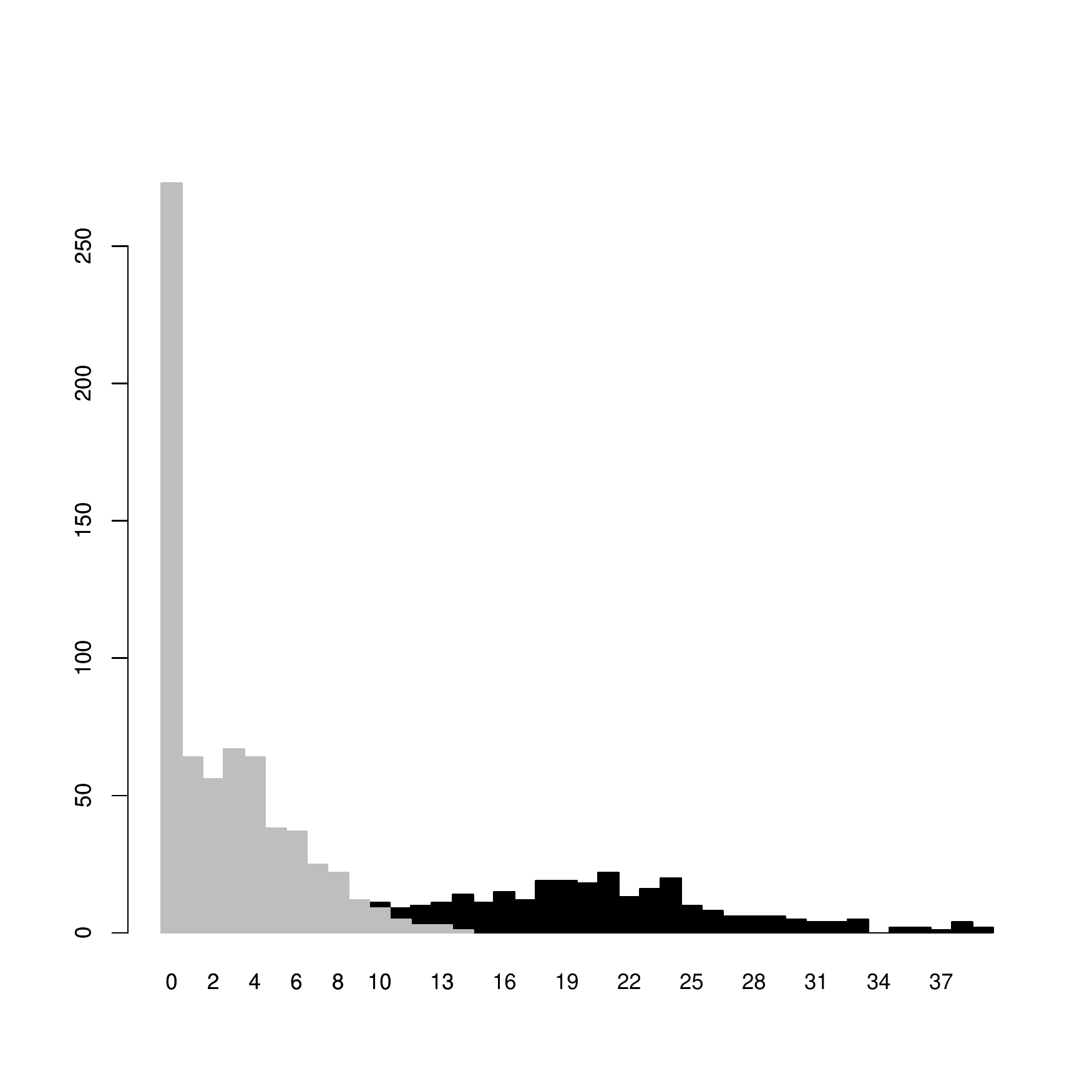}\\
			\textbf{ZIGP}$_{10}$\\
			ZIGP($\eta = 0.40, \lambda = 3, \theta=0.30$)\\
		\end{tabular}
		& \begin{tabular}{c}
			\includegraphics[width=0.3\linewidth]{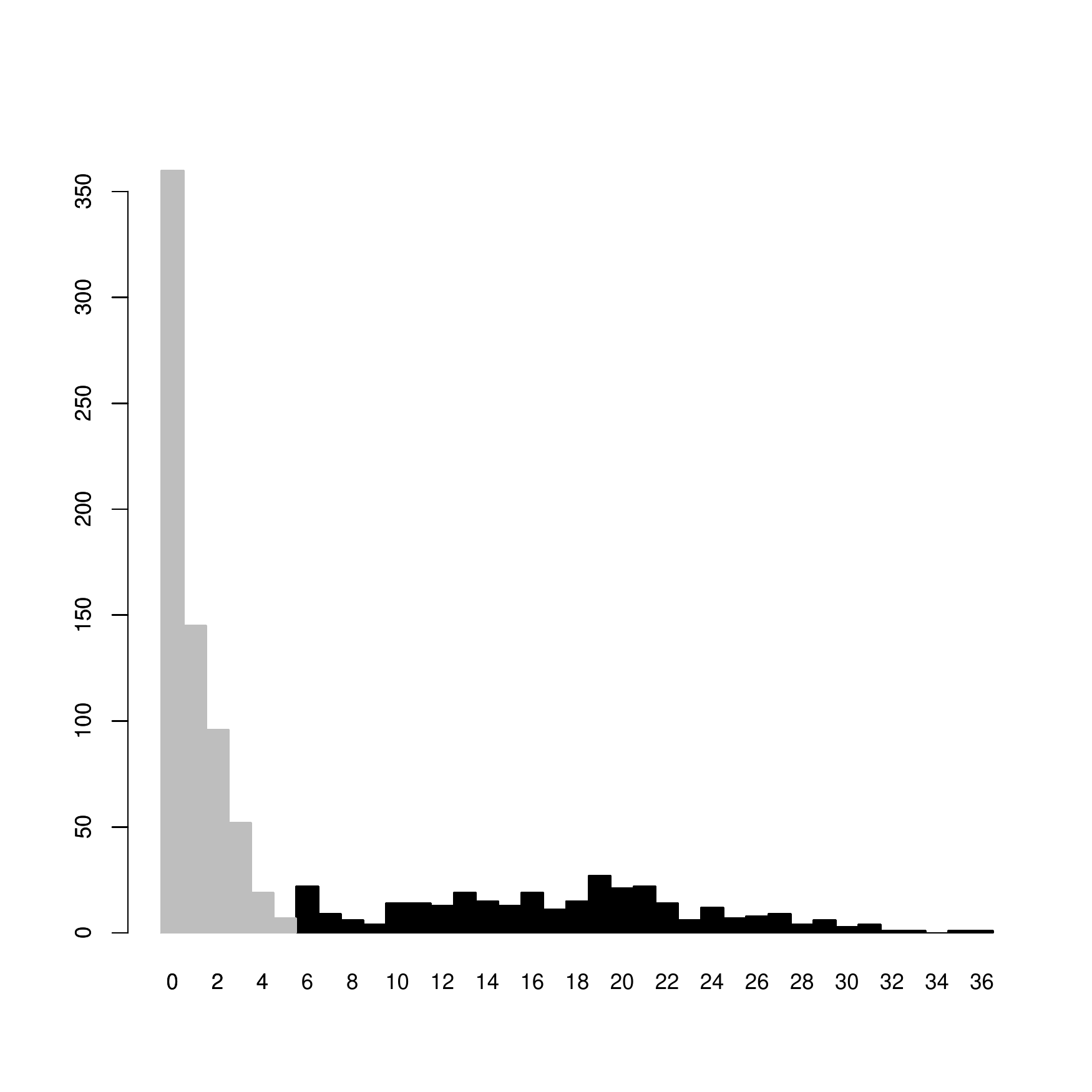}\\
			\textbf{ZIP}$_5$\\
			ZIP($\eta = 0.40, \lambda = 1.5$)\\
		\end{tabular}\\
	\end{tabular}
\end{center}

\vspace{10mm}
\begin{center}
	\textbf{Table 6.}  Numerical Comparison using $C_1$ as cut-off when the Non-null Distribution is Binomial($n=250, p=0.20$), $\pi_0 = 0.70$ and $\alpha = 0.05$.  The number in $(\cdot)$ represents the standard deviation.\\
	\vspace{3mm}
	\begin{tabular}{ll|ccc|ccc|ccc}
		\hline
		\hline
		& & \multicolumn{3}{c|}{Two-Stage Procedure} & \multicolumn{3}{c|}{One-Stage Procedure} &
		\multicolumn{3}{c}{Storey's FDR} \\
		\hline
		Null & Model & & & &  & & &\\
		Distribution &  for $f_0$ & $R$ & $\widehat{FDR}$ & $\widehat{TPR}$ & $R$ & $\widehat{FDR}$ & $\widehat{TPR}$ & $R$ & $\widehat{FDR}$ & $\widehat{TPR}$ \\
		\hline
		&& &&& &&& &&\\
		ZIGP$_9$ & 	ZIGP & 316.41 & 0.05152 & 0.99987 & 283.86 & 0.01551 & 0.93132 & 288.81 & 0.02115 & 0.94208 \\
		& & (15.24) & (0.02046) & (0.00248) & (14.30) & (0.00882) & (0.01549) & (14.41) & (0.00994) & (0.01644) \\
		&ZIP & 311.60 & 0.04190 & 0.99472 & 311.44 & 0.04171 & 0.99472 & 307.52 & 0.03854 & 0.98507 \\
		& & (15.96) & (0.01461) & (0.01495) & (16.15) & (0.01489) & (0.01495) & (17.17) & (0.01521) & (0.02425) \\
		&GP & 1.22 & 0.00000 & 0.00408 & 0.00 &  & 0.00180 & 4.01 & 0.00000 & 0.01335 \\
		& & (0.55) & (0.00000) & (0.00184) & (0.00) & (0.00000) & (0.00221) & (2.40) & (0.00000) & (0.00796) \\
		&P & 377.83 & 0.20463 & 1.00000 & 377.83 & 0.20463 & 1.00000 & 345.32 & 0.13101 & 1.00000 \\
		& & (20.84) & (0.03695) & (0.00000) & (20.84) & (0.03695) & (0.00000) & (14.79) & (0.01855) & (0.00000) \\
		&& &&& &&& &&\\
		\hline
		&& &&& &&& &&\\
		ZIGP$_{10}$ &ZIGP& 237.25 & 0.04893 & 0.74353 & 230.78 & 0.04185 & 0.74340 & 248.22 & 0.04711 & 0.78263 \\
		& & (90.17) & (0.03985) & (0.27104) & (87.49) & (0.03263) & (0.27096) & (67.64) & (0.03405) & (0.19811) \\
		&ZIP & 375.04 & 0.22727 & 0.96514 & 328.08 & 0.13961 & 0.94225 & 320.20 & 0.12648 & 0.92887 \\
		& & (19.83) & (0.02554) & (0.01382) & (39.07) & (0.06537) & (0.02774) & (27.39) & (0.04568) & (0.02494) \\
		&GP & 1.20 & 0.00000 & 0.00399 & 0.00 & 0.00000 & 0.00399 & 10.30 & 0.00000 & 0.03430 \\
		& & (0.50) & (0.00000) & (0.00165) & (0.00) & (0.00000) & (0.00165) & (5.04) & (0.00000) & (0.01658) \\
		&P & 611.10 & 0.50773 & 1.00000 & 611.10 & 0.50773 & 1.00000 & 577.55 & 0.47925 & 1.00000 \\
		& & (34.36) & (0.03149) & (0.00000) & (34.36) & (0.03149) & (0.00000) & (31.39) & (0.03244) & (0.00000) \\
		&& &&& &&& &&\\
		\hline
		&& &&& &&& &&\\
		ZIP$_5$ & 	ZIGP & 301.86 & 0.00596 & 1.00000 & 279.17 & 0.00021 & 0.93021 & 283.37 & 0.00082 & 0.94361 \\
		& & (14.07) & (0.00489) & (0.00000) & (13.54) & (0.00090) & (0.01482) & (14.02) & (0.00179) & (0.01624) \\
		&ZIP & 287.81 & 0.00186 & 0.95733 & 285.50 & 0.00150 & 0.95699 & 287.33 & 0.00171 & 0.95592 \\
		& & (14.89) & (0.00300) & (0.01938) & (15.47) & (0.00293) & (0.01939) & (14.75) & (0.00275) & (0.02013) \\
		&GP & 258.70 & 0.00370 & 0.86037 & 242.07 & 0.00015 & 0.80723 & 248.86 & 0.00042 & 0.83079 \\
		& & (88.48) & (0.00471) & (0.29058) & (81.70) & (0.00074) & (0.26859) & (79.44) & (0.00125) & (0.26233) \\
		&P & 338.33 & 0.10956 & 1.00000 & 338.33 & 0.10956 & 1.00000 & 322.21 & 0.06800 & 1.00000 \\
		& & (26.65) & (0.05600) & (0.00000) & (26.65) & (0.05600) & (0.00000) & (16.92) & (0.02948) & (0.00000) \\
		&& &&& &&& &&\\
		\hline
		\hline
	\end{tabular}
\end{center}

\newpage
\begin{center}
	\textbf{Table 7.}  Numerical Comparison using $C_2$ as cut-off when the Non-null Distribution is Binomial($n=250, p=0.20$), $\pi_0 = 0.80$ and $\alpha = 0.05$.  The number in $(\cdot)$ represents the standard deviation.\\
	\vspace{3mm}
	\begin{tabular}{ll|ccc|ccc|ccc}
		\hline
		\hline
		& & \multicolumn{3}{c|}{Two-Stage Procedure} & \multicolumn{3}{c|}{One-Stage Procedure} &
		\multicolumn{3}{c}{Storey's FDR} \\
		\hline
		Null & Model & & & &  & & &\\
		Distribution &  for $f_0$ & $R$ & $\widehat{FDR}$ & $\widehat{TPR}$ & $R$ & $\widehat{FDR}$ & $\widehat{TPR}$ & $R$ & $\widehat{FDR}$ & $\widehat{TPR}$ \\
		\hline
		&& &&& &&& &&\\
		ZIGP$_{9}$ & 	ZIGP & 314.70 & 0.04721 & 0.99907 & 282.66 & 0.01445 & 0.92840 & 286.89 & 0.01901 & 0.93786 \\
		& & (15.04) & (0.01780) & (0.00898) & (14.18) & (0.00794) & (0.01745) & (14.61) & (0.00978) & (0.01796) \\
		&ZIP & 312.34 & 0.04253 & 0.99648 & 312.23 & 0.04239 & 0.99648 & 309.36 & 0.03991 & 0.98951 \\
		& & (15.43) & (0.01447) & (0.01225) & (15.58) & (0.01468) & (0.01225) & (16.99) & (0.01486) & (0.02147) \\
		&GP & 1.22 & 0.00000 & 0.00408 & 0.00 & 0.00000 & 0.00041 & 2.23 & 0.00000 & 0.00741 \\
		& & (0.55) & (0.00000) & (0.00184) & (0.00) & (0.00000) & (0.00123) & (1.52) & (0.00000) & (0.00504) \\
		&P & 377.83 & 0.20463 & 1.00000 & 377.83 & 0.20463 & 1.00000 & 345.32 & 0.13101 & 1.00000 \\
		& & (20.84) & (0.03695) & (0.00000) & (20.84) & (0.03695) & (0.00000) & (14.79) & (0.01855) & (0.00000) \\
		&& &&& &&& &&\\
		\hline
		&& &&& &&& &&\\
		ZIGP$_{10}$ &ZIGP& 97.34 & 0.02279 & 0.30289 & 92.95 & 0.02126 & 0.30282 & 139.75 & 0.02109 & 0.44630 \\
		& & (130.74) & (0.03934) & (0.40247) & (127.11) & (0.03328) & (0.40237) & (100.95) & (0.03428) & (0.30743) \\
		&ZIP & 375.16 & 0.22740 & 0.96525 & 334.29 & 0.15174 & 0.94528 & 324.19 & 0.13421 & 0.93174 \\
		& & (20.02) & (0.02568) & (0.01387) & (40.71) & (0.06618) & (0.03121) & (28.79) & (0.04682) & (0.02613) \\
		&GP & 1.20 & 0.00000 & 0.00399 & 0.00 & 0.00000 & 0.00399 & 11.26 & 0.00000 & 0.03751 \\
		& & (0.50) & (0.00000) & (0.00165) & (0.00) & (0.00000) & (0.00165) & (4.89) & (0.00000) & (0.01598) \\
		&P & 611.10 & 0.50773 & 1.00000 & 611.10 & 0.50773 & 1.00000 & 577.55 & 0.47925 & 1.00000 \\
		& & (34.36) & (0.03149) & (0.00000) & (34.36) & (0.03149) & (0.00000) & (31.39) & (0.03244) & (0.00000) \\
		&& &&& &&& &&\\
		\hline
		&& &&& &&& &&\\
		ZIP$_{5}$ & 	ZIGP & 301.12 & 0.00611 & 0.99739 & 275.43 & 0.00011 & 0.91829 & 279.84 & 0.00028 & 0.93237 \\
		& & (14.39) & (0.00556) & (0.01144) & (14.21) & (0.00064) & (0.01818) & (13.75) & (0.00102) & (0.01573) \\
		&ZIP & 288.73 & 0.00219 & 0.96005 & 285.97 & 0.00176 & 0.95792 & 287.75 & 0.00193 & 0.95708 \\
		& & (15.34) & (0.00343) & (0.02126) & (16.18) & (0.00340) & (0.02223) & (15.28) & (0.00314) & (0.02300) \\
		&GP & 221.18 & 0.00372 & 0.73562 & 204.70 & 0.00018 & 0.68316 & 209.69 & 0.00042 & 0.70059 \\
		& & (128.96) & (0.00472) & (0.42671) & (119.76) & (0.00081) & (0.39767) & (119.83) & (0.00125) & (0.39860) \\
		&P & 338.33 & 0.10956 & 1.00000 & 338.33 & 0.10956 & 1.00000 & 322.21 & 0.06800 & 1.00000 \\
		& & (26.65) & (0.05600) & (0.00000) & (26.65) & (0.05600) & (0.00000) & (16.92) & (0.02948) & (0.00000) \\
		&& &&& &&& &&\\
		\hline
		\hline
	\end{tabular}
\end{center}
\newpage

\end{document}